\documentclass[12pt]{article}
\usepackage{graphicx}

\newcommand\pubdate{July 12, 2011}

\newcommand\pubnumber{FERMILAB-CONF-11-317-E}

\def\Title#1{\begin{center} {\Large #1 } \end{center}}
\def\Author#1{\begin{center}{ \sc #1} \end{center}}
\def\Address#1{\begin{center}{ \it #1} \end{center}}

\newcommand\pubblock{\rightline{\begin{tabular}{l} \pubnumber\\
         \pubdate  \end{tabular}}}
\newenvironment{Abstract}{\begin{center}{\bf Abstract}\end{center} \bigskip \begin{quotation}  }{\end{quotation}}
\newenvironment{Presented}{\begin{quotation} \begin{center} 
             PRESENTED AT\end{center}\bigskip 
      \begin{center}\begin{large}}{\end{large}\end{center} \end{quotation}}
\def\Acknowledgements{\bigskip  \bigskip \begin{center} \begin{large}
             \bf ACKNOWLEDGEMENTS \end{large}\end{center}}

\def\beq{\begin{equation}}
\def\eeq#1{\label{#1}\end{equation}}
\def\eeqn{\end{equation}}

\def\beqa{\begin{eqnarray}}
\def\eeqa#1{\label{#1}\end{eqnarray}}
\def\eeqan{\end{eqnarray}}

\let\bar=\overbar

\def\Dslash{\not{\hbox{\kern-4pt $D$}}}
\def\dslash{\not{\hbox{\kern-2pt $\del$}}}

\def\msb{{\bar{\ssstyle M \kern -1pt S}}}

\newcommand{\chiminus}{\ensuremath{\tilde{\chi^-_1}}}

\newcommand{\chipm}{\ensuremath{\tilde{\chi^{\pm}_1}}}
\newcommand{\chineu}{\ensuremath{\tilde{\chi^0_1}}}
\newcommand{\chineuhi}{\ensuremath{\tilde{\chi^0_2}}}
\newcommand{\sbot}{\ensuremath{\tilde{b_1}}}
\newcommand{\sctop}{\ensuremath{\tilde{t_1}}}
\newcommand{\sneu}{\ensuremath{\tilde{\nu}}}
\newcommand{\epm}{\ensuremath{e^{\pm}}}

\newcommand{\mump}{\ensuremath{\mu^{\mp}}}
\newcommand{\grav}{\ensuremath{\tilde{G}}}

\def\slashchar#1{\setbox0=\hbox{$#1$}           
  \dimen0=\wd0                                 
  \setbox1=\hbox{/} \dimen1=\wd1               
  \ifdim\dimen0>\dimen1                        
     \rlap{\hbox to \dimen0{\hfil/\hfil}}      
     #1                                        
  \else                                        
     \rlap{\hbox to \dimen1{\hfil$#1$\hfil}}   
     /                                         
  \fi}                                         %
\def\met{\slashchar{E}_T}                    %
\def\slashchar#1{\setbox0=\hbox{$#1$}           
  \dimen0=\wd0                                 
  \setbox1=\hbox{/} \dimen1=\wd1               
  \ifdim\dimen0>\dimen1                        
     \rlap{\hbox to \dimen0{\hfil/\hfil}}      
     #1                                        
  \else                                        
     \rlap{\hbox to \dimen1{\hfil$#1$\hfil}}   
     /                                         
  \fi}                                         %
\def\slashchar#1{\setbox0=\hbox{$#1$}           
  \dimen0=\wd0                                 
  \setbox1=\hbox{/} \dimen1=\wd1               
  \ifdim\dimen0>\dimen1                        
     \rlap{\hbox to \dimen0{\hfil/\hfil}}      
     #1                                        
  \else                                        
     \rlap{\hbox to \dimen1{\hfil$#1$\hfil}}   
     /                                         
  \fi}                                         %
\begin{document}
\begin{titlepage}
\pubblock

\vfill


\Title{SUSY Searches at the Tevatron}
\vfill
\Author{L.~Zivkovic}  
\Address{Brown University, Providence, RI 02912, USA}
\vfill


\begin{Abstract}
In this article results from supersymmetry searches at D0 and CDF are reported. 
Searches  for third generation squarks,  searches for gauginos,
and searches for models with R-parity violation are described.
As no signs of supersymmetry for these models are observed, the most stringent limits to date are presented.
\end{Abstract}

\vfill

\begin{Presented}
The Ninth International Conference on\\
Flavor Physics and CP Violation\\
(FPCP 2011)\\
Maale Hachamisha, Israel,  May 23--27, 2011
\end{Presented}
\vfill

\end{titlepage}
\def\thefootnote{\fnsymbol{footnote}}
\setcounter{footnote}{0}
%


\section{Introduction}

The standard model (SM) describes matter and its interactions at the elementary level. However it is not a complete theory, 
since there are indications of phenomena
which the SM does not describe. Supersymmetry (SUSY)~\cite{Martin:1997ns} is one of the extensions of the SM that 
provides a solution to many shortcomings of the SM. SUSY is the 
symmetry between different states of spin, i.e. between fermions and bosons. 
Since each SM particle has its SUSY partner, sparticle, the number of parameters in SUSY is 
increased. The minimal supersymmetric standard model (MSSM) is a model with minimal particle content. It is possible in the MSSM that the third generation 
squarks are lighter than the first and second, thus making them accessible to Tevatron experiments.

It is known that particles and sparticles do not have the same mass, which is an indication that SUSY is broken. There are several possible mechanisms of 
SUSY breaking. D0 and CDF have searched for gauginos within the minimal supergravity model (mSUGRA), and gauge mediated symmetry breaking model (GMSB).
The mSUGRA is described by five independent parameters: the unified scalar and gaugino masses $m_0$ and $m_{1/2}$, the ratio of the 
vacuum expectation values of the two Higgs doublets $\tan\beta$, the unified trilinear coupling $A_0$, and the sign of the Higgs mass parameter sign$(\mu)$.
GMSB is described by six parameters: the breaking scale $\Lambda$, the mass and number of messengers $M$ and $N$, the scale factor of gravitino mass 
$C_{grav}$, $\tan\beta$ and the sign$(\mu)$.

R-parity, defined as $R=(-1)^{3(B-L)+2S}$, where $B$ is baryon number, $L$ is lepton number, and $S$ is spin,  is introduced to preserve 
lepton and baryon number conservation. If it is conserved the lightest supersymmetric particle (LSP) is stable, all sparticles eventually decay to LSP, and sparticles are 
produced in pairs. However, lepton and baryon number symmetry is accidental, so R-parity is imposed by hand. If R-parity is violated SUSY particles decay to SM 
particles and they can be produced singly, which leads to higher cross sections at Tevatron.

The CDF~\cite{CDFdet} and D0~\cite{Abazov:2005pn, Abolins:2007yz, Angstadt:2009ie} detectors 
are multi purpose detectors. They consist of central tracking system, electromagnetic and hadronic calorimeters and 
outer muon detectors. They recorded about 10~fb$^{-1}$ of data between April 2002 and May 2011. Results presented in this article use up to 
6.3~fb$^{-1}$.

\section{Third generation squark searches}

\begin{figure}[htb]
\centering
\includegraphics[width=0.45\textwidth]{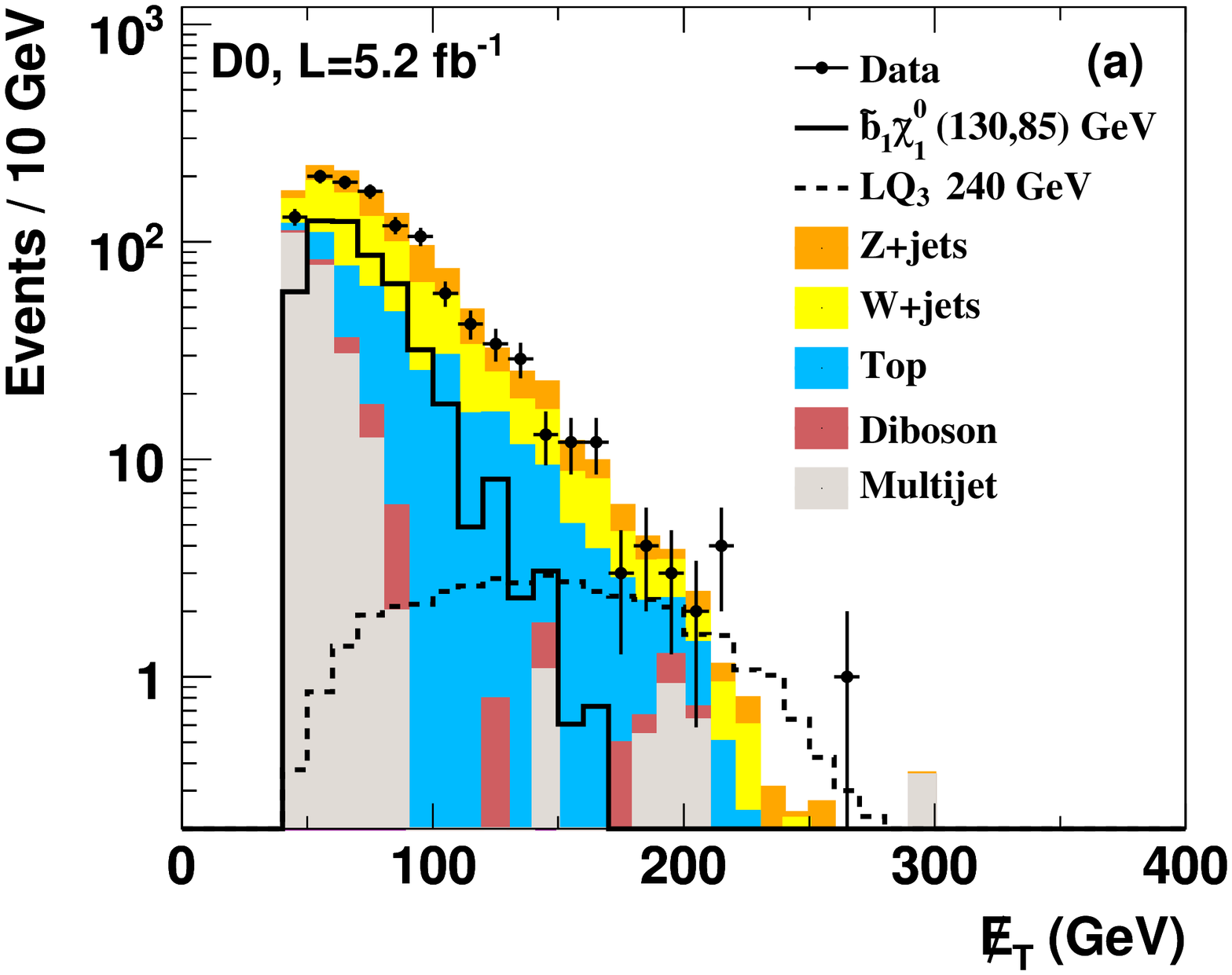}
\includegraphics[width=0.45\textwidth]{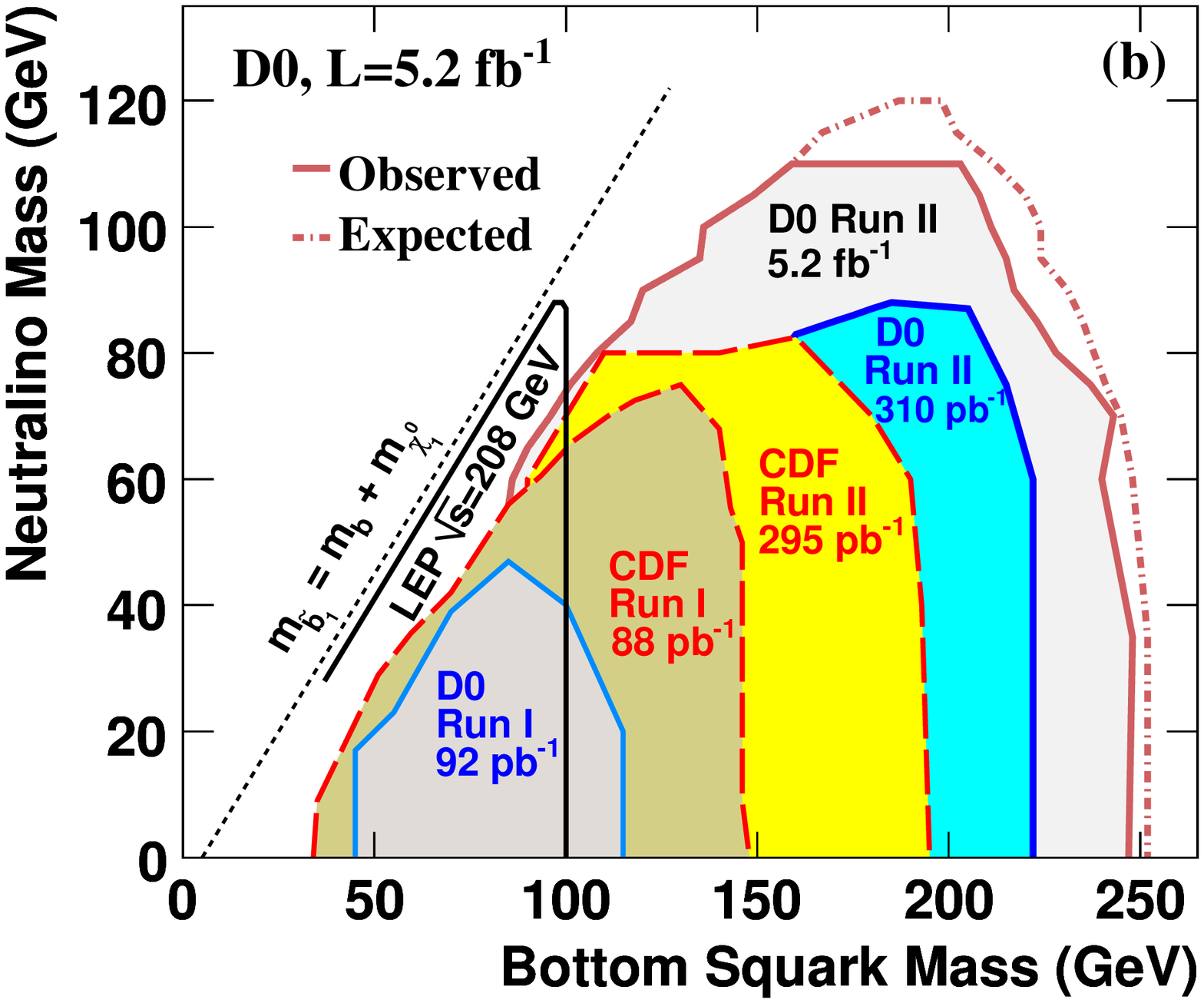}
\caption{The $\met$ distribution in the $\sbot\bar{\sbot}\rightarrow b\chineu\bar{b}\bar{\chineu}$ search for data, expected SM backgrounds, and sbottom quark signal for $m_{\sbot}=130$~GeV and 
$m_{\chineu}=85$~GeV (left).
The observed and expected 95~\%~C.L. exclusion contour in a $m_{\sbot}$ and $m_{\chineu}$ plane (right).}
\label{fig:sbot-met}
\end{figure}

The D0 experiment searched for pair production of sbottom quarks~\cite{Abazov:2010wq} in 5.2~fb$^{-1}$ of data.
It was assumed that the mass of the sbottom quark satisfied $m_b+m_{\chineu}<m_{\sbot}<m_t+m_{\chiminus}$,
so only $\sbot\rightarrow b\chineu$ was possible, and that $\chineu$ was the LSP. The search was interpreted in the MSSM with R-parity conservation.
The signature of this final state was two $b$ jets and large missing transverse energy $\met$, so data were recorded using a 
combination of jet and $\met$ based triggers. 
The dominant SM background was from multijet (MJ) events where instrumental $\met$ was present due to jet resolution and mismeasurement.
It was estimated from data. Other backgrounds included $W/Z$+jets, $t\bar{t}$ and diboson events, which were obtained from Monte Carlo (MC) simulation.
Events were required to have at least two not back-to-back $b$-tagged jets with $p_T>20$~GeV and $\met>40$~GeV.
Selection was further optimized for different choices of ($m_{\sbot},m_{\chineu}$) to achieve the maximal sensitivity.
Good agreement between data and SM backgrounds was observed as can be seen in Fig.~\ref{fig:sbot-met} (left) and Table~\ref{tab:d0-sbot}.
A 95\% C.L limit on the production of sbottom quark pairs was set and the excluded region is shown in the plane of
the bottom squark versus neutralino mass in Fig.~\ref{fig:sbot-met} (right).
\begin{table}[!hbtp]
\begin{center}
\begin{tabular}{l|cc}  
\hline\hline
  $m_{\sbot},m_{\chineu}$ &  (130,85) &  (240,0) \\ \hline
 Data     &  901  &    7 \\
 Background     &  $971\pm 52$   &     $6.9 \pm 1.7$  \\
  Signal     &  $481\pm 66$   &     $10.5 \pm 1.9$   \\
 
\hline\hline
\end{tabular}
\caption{Number of events for the two chosen $(m_{\sbot},m_{\chineu})$ points for data, SM expectation and sbottom quark signal.}
\label{tab:d0-sbot}
\end{center}
\end{table}

\begin{figure}[htb]
\centering
\includegraphics[width=0.45\textwidth]{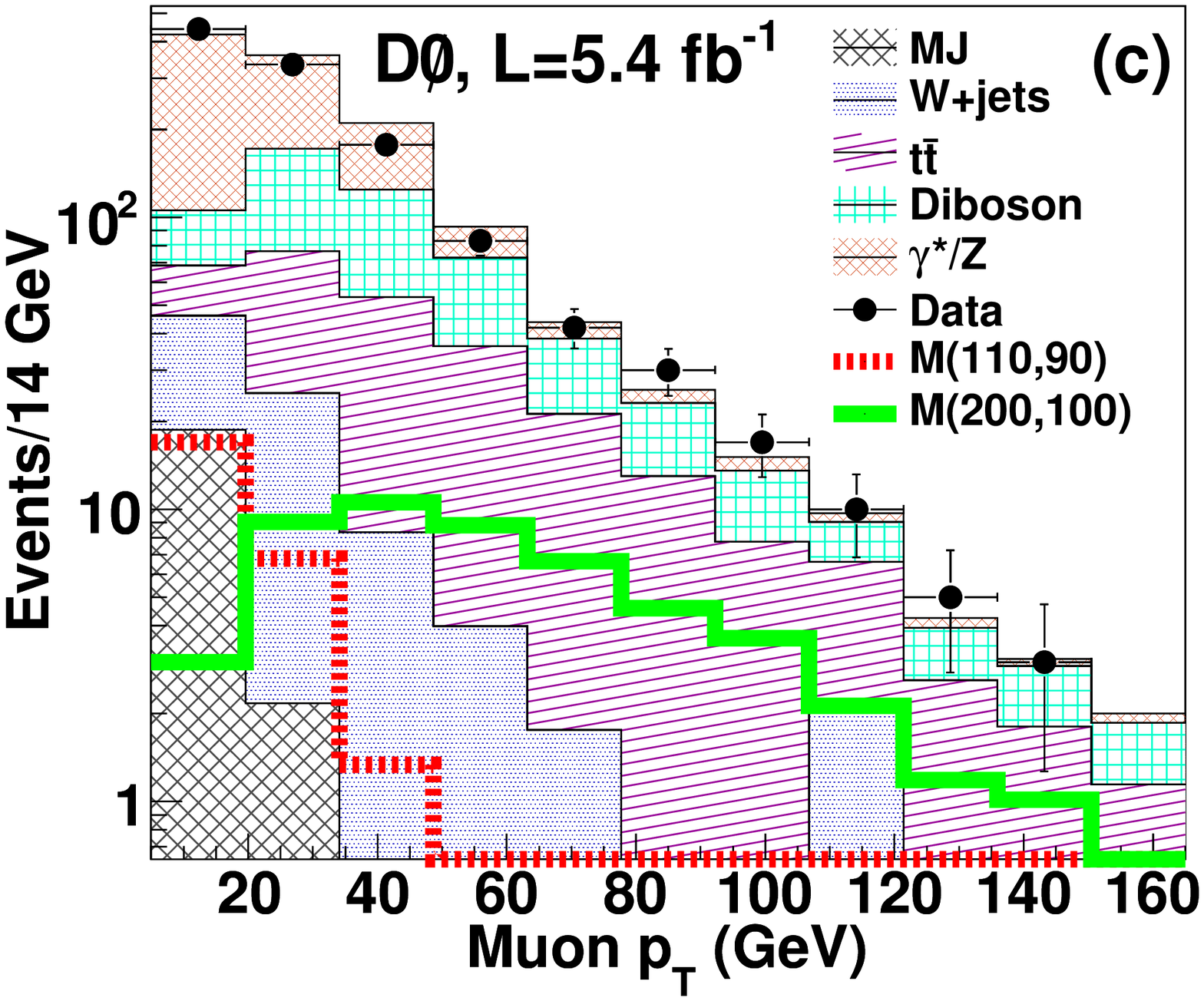}
\includegraphics[width=0.45\textwidth]{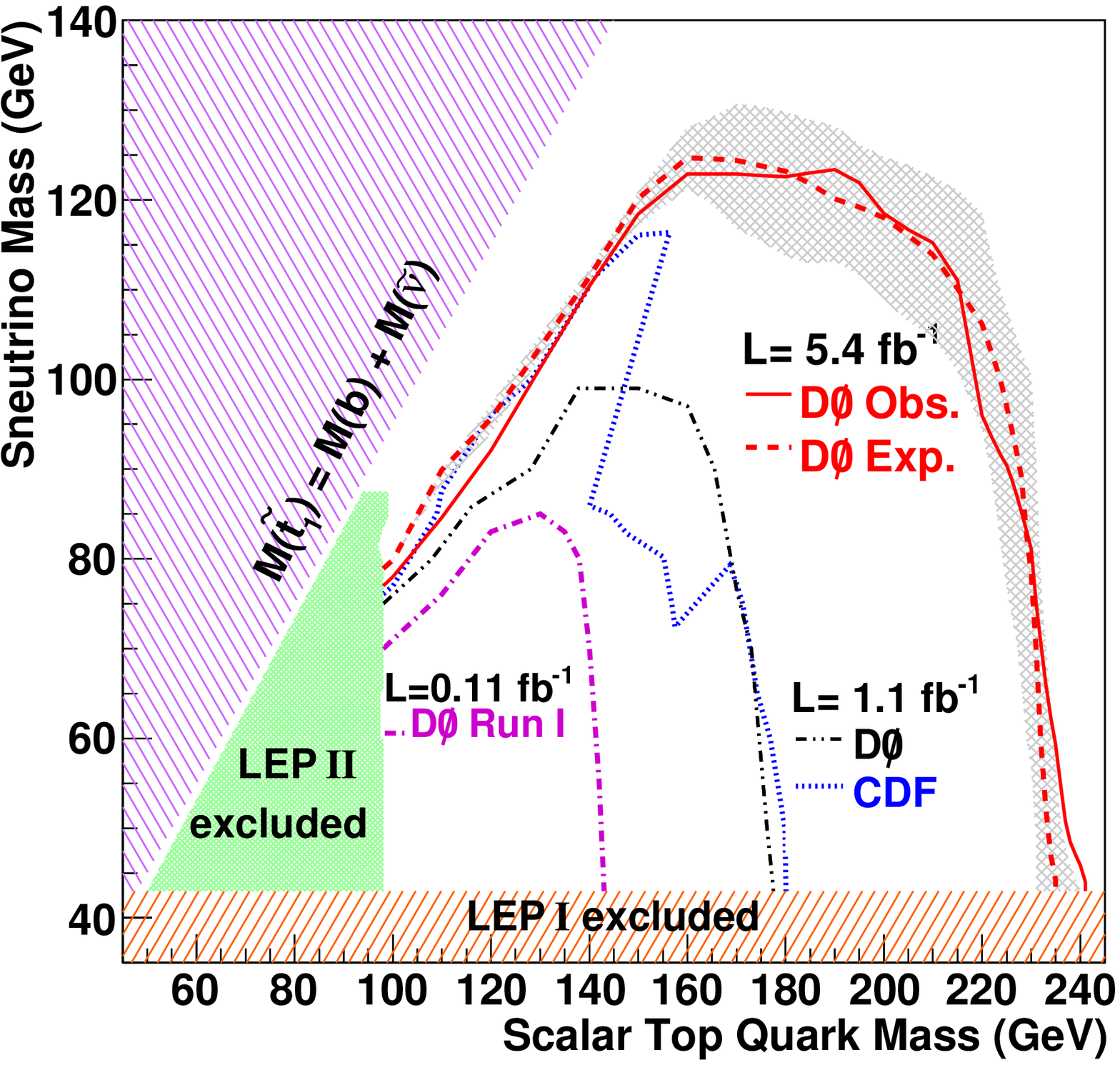}
\caption{The $p_T$ of the muon for data, expected background and two different $M_{\sctop}$ representing small$-\Delta M$ and large$-\Delta M$ (left). 
The observed and expected 95\% C.L. exclusion contours in the $(m_{\sctop},m_{\sneu})$ plane (right).}
\label{fig:d0-stop}
\end{figure}

D0~\cite{Abazov:2010xm} and CDF~\cite{Aaltonen:2009sf} searched for pair production of scalar top quarks. 
D0 searched for stop quark pair production $\sctop\bar{\sctop}$ with the
$b\bar{b}\epm\mump\sneu\bar{\sneu}$ final state in 5.4~fb$^{-1}$ of data. It was assumed that the branching ratio for $\sctop\rightarrow bl\sneu$ was 100\%
and that the sneutrino $\sneu$ was either the LSP or that it decayed invisibly into a neutrino and a neutralino. 
Events were selected with one isolated electron with $p_T>15$~GeV and 
$|\eta|<1.1$~\cite{eta}, one isolated muon with $p_T>10$~GeV and $|\eta|<2$, and $\met>7$~GeV at the preselection level.
The dominant SM backgrounds included 
$Z\rightarrow\tau\tau$ where both $\tau$ leptons decayed leptonically, top quark pairs, diboson production, $W+$jets and 
instrumental MJ background. To further optimize this search, samples were divided according to the mass difference between $\sctop$ and $\sneu$, 
$\Delta M=M_{\sctop}-M_{\sneu}$. D0 chose two benchmark points, $(M_{\sctop},M_{\sneu})=(200,100)$~GeV for "large-$\Delta M$", 
$\Delta M>60$~GeV, and $(M_{\sctop},M_{\sneu})=(110,90)$~GeV for "small-$\Delta M$", $\Delta M<60$~GeV. 
Signal selection was optimized as a function of $\Delta M$.
In a dominant background after preselection, $Z\rightarrow\tau\tau$, two leptons were often back to back, and $\met$ was low. 
Thus, an additional requirement was 
introduced, events were rejected if $\Delta\phi(e,\mu)>2.8$ and $\met<20$~GeV.
Fig.~\ref{fig:d0-stop} (left) shows muon $p_T$ after this selection.
To discriminate against dominant backgrounds ($Z\rightarrow\tau\tau$, $t\bar{t}$ and diboson) and signal, a separate multivariate analysis output (MVA)
was built for each background.
A cut was then applied on the most discriminating MVA, which was $t\bar{t}$ for "small-$\Delta M$", and  $Z\rightarrow\tau\tau$ for 
"large-$\Delta M$". A two dimensional histogram of the remaining two MVA distributions was used to search for signal. 
In the absence of any significant excess, the 95\% 
C.L. exclusion limits on scalar top quark production as a function
of the $\sneu$ and $\sctop$ masses was set, and is shown in Fig.~\ref{fig:d0-stop} (right).  

\begin{figure}[htb]
\centering
\includegraphics[width=0.45\textwidth]{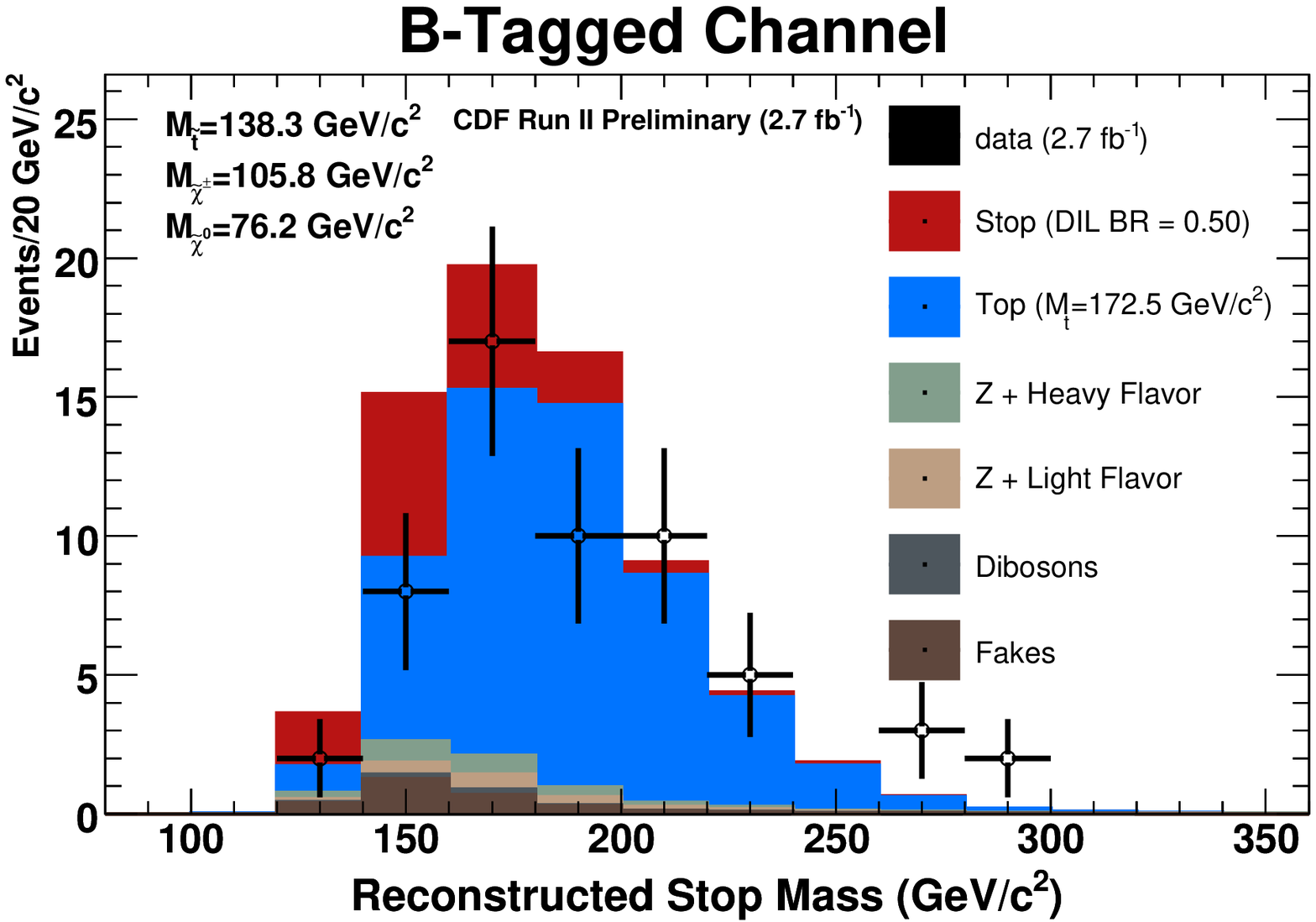}
\includegraphics[width=0.45\textwidth]{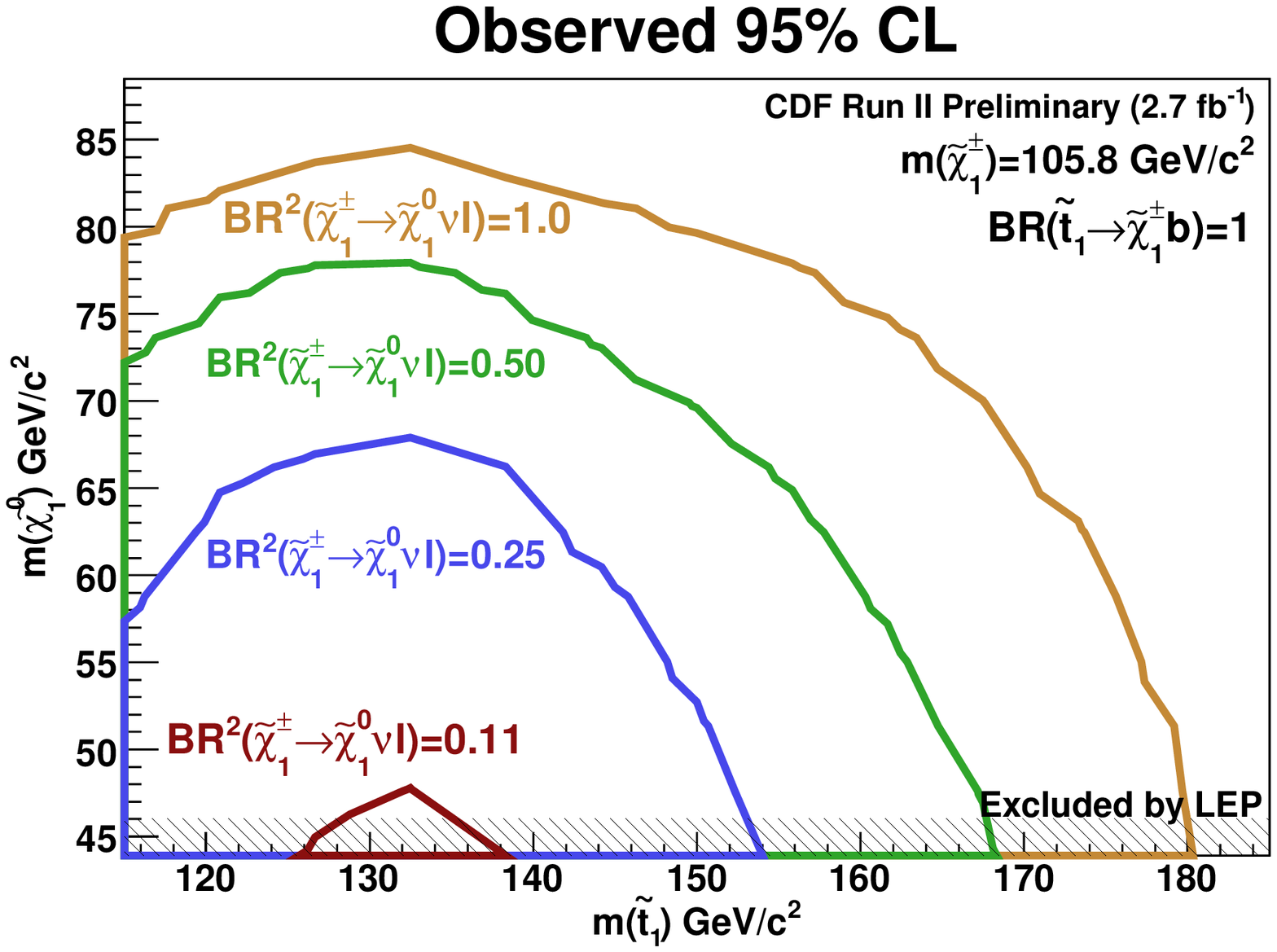}
\caption{Reconstructed $m_{\sctop}$ for data, predicted background and expected signal for events with at least one $b$-tagged jet  (left). The observed 
95\% C.L. limits in $(m_{\sctop},m_{\chineu})$ plane for various branching ratios to the dilepton final state (right). }
\label{fig:cdf-stop}
\end{figure}

CDF searched for the pair production of scalar top quarks further decaying to the final state with two leptons,  
$\sctop\rightarrow b\chipm\rightarrow b\chineu l\nu$ in 2.7~fb$^{-1}$ of data. It was assumed $m_{\sctop} < m_t$,  neutralino $\chineu$ is LSP, 
$m_{\chipm}<m_{\sctop}-m_b$
and R-parity conservation.
Depending on the masses of chargino, $\chipm$, and neutralino $\chineu$, and chosen SUSY parameters, the branching ratio 
$BR(\chipm\rightarrow b\chineu l\nu)$ can have values between 0.11 and 1.0.
The final state, containing two leptons, two $b$ quarks and $\met$ from neutralino is identical to top quark pair production. 
Thus the reconstructed top mass will be biased
toward lower values in the presence of scalar top quarks.
Events were selected with two leptons with $p_T>20$~GeV, two jets with $p_T>12$~GeV, and $\met>20$~GeV. The dominant SM backgrounds were $t\bar{t}$, 
$Z+jets$, diboson and instrumental MJ background. To maximize sensitivity of this search, samples were divided into events with no-$b$-tagged jets 
and at least one $b$-tagged jet. Data was in agreement with SM prediction as shown in 
Table~\ref{tab:cdf-stop}. Scalar top yield was calculated for $m_{\sctop}=132.5$~GeV, 
$m_{\chipm}=105.8$~GeV, $m_{\chineu}=47.6$~GeV and $BR=0.11$. Figure~\ref{fig:cdf-stop} (left) 
shows the reconstructed stop quark mass for $m_{\sctop}=138.3$~GeV, 
$m_{\chipm}=105.8$~GeV, $m_{\chineu}=76.2$~GeV and $BR=0.5$. In the absence of a significant excess in data mass limits in
the $(m_{\chineu},m_{\sctop})$ 
plane for different branching ratios was set (see Fig.~\ref{fig:cdf-stop} (right)).

\begin{table}[!hbtp]
\begin{center}
\begin{tabular}{l|ccc}  
\hline\hline
   &  Data &  Total MC & Signal\\ \hline
 No $b$-tag     &  65  &     $65.9 \pm 9.8$ & $3.9 \pm 0.9$ \\
 $\geq 1 b$-tag     &  57   &     $56.4 \pm 7.2$ & $9.5 \pm 1.9$  \\
 
\hline\hline
\end{tabular}
\caption{Event yields for data, predicted backgrounds and expected signal with no $b$-tagged jets and with at least one $b$-tagged jet.}
\label{tab:cdf-stop}
\end{center}
\end{table}

\section{Gaugino Searches}

\begin{figure}[htb]
\centering
\includegraphics[width=0.45\textwidth]{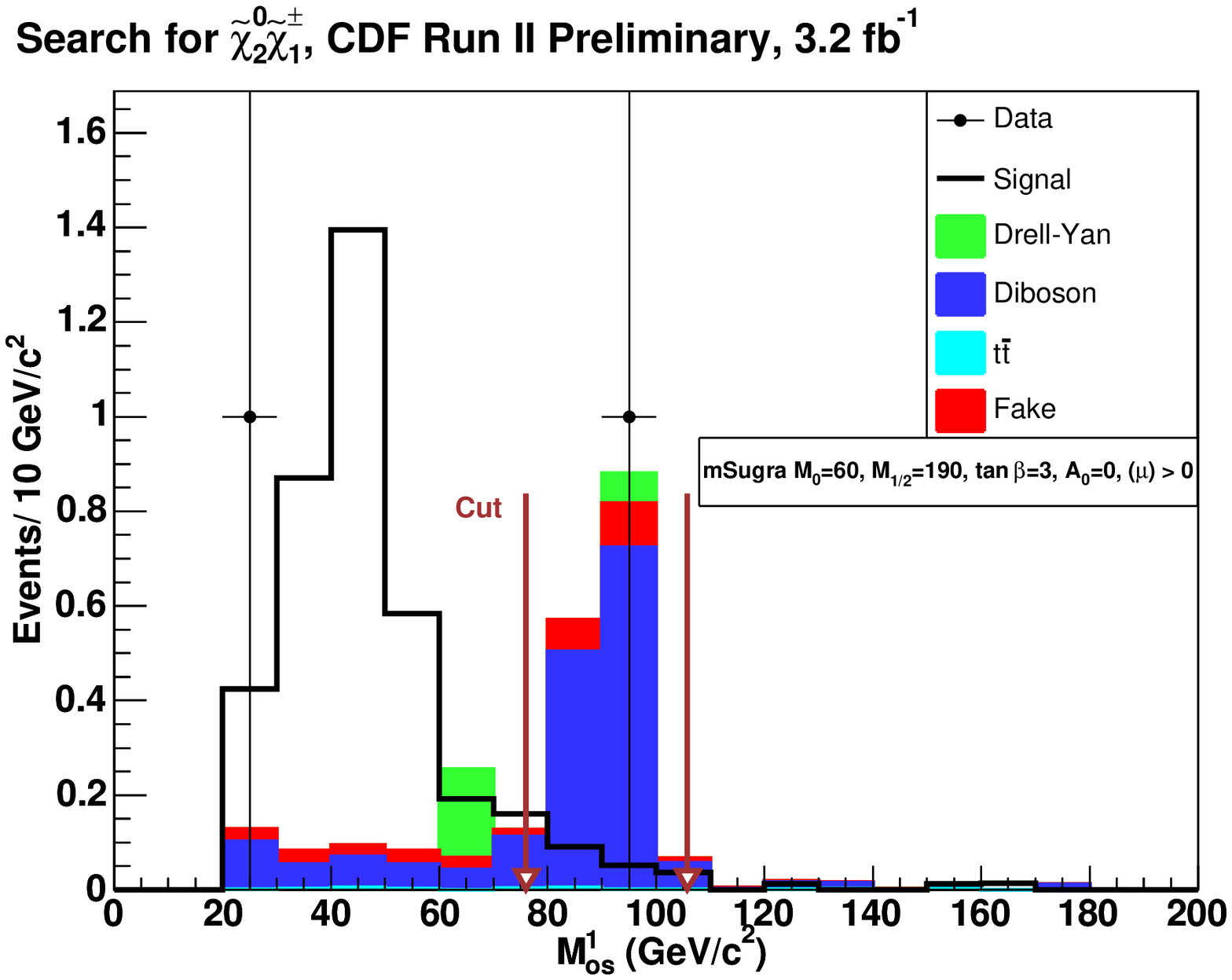}
\includegraphics[width=0.45\textwidth]{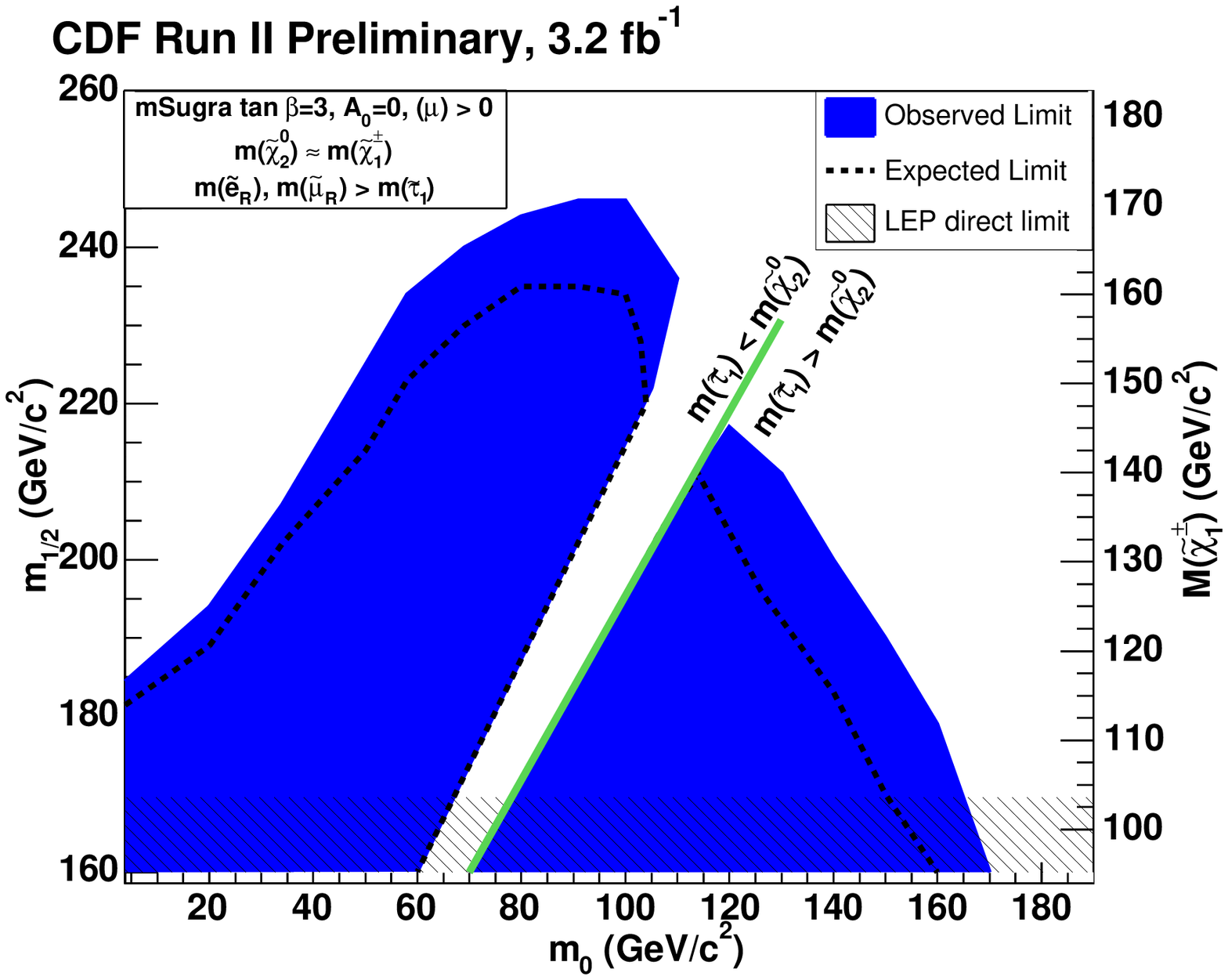}
\caption{Invariant mass of the two opposite sign leptons after preselection cuts (left). Expected and observed 95\% C.L. limit contours for mSUGRA in 
a $(m_0,m_{1/2})$ plane (right). }
\label{fig:cdf-trilep}
\end{figure}

CDF~\cite{Aaltonen:new} and D0~\cite{Abazov:2009zi} searched for the production of charginos and neutralinos in the final 
states with at least three leptons in 3.2~fb$^{-1}$ (CDF) and 2.3~fb$^{-1}$ (D0) of data.
At CDF events were divided in search channels based on a lepton flavor and quality. Electrons and muons can be tight or loose, and tracks were always tight.
Each event was categorized into an exclusive trilepton channel composed of combinations of these objects.
All objects were required to be central, $|\eta|<1$, and isolated from other objects. Opposite sign leptons were required not to
be back to back, and their invariant mass to be outside the $Z$ mass window. In addition events were selected if they have less than two jets. 
The invariant mass after all
cuts except the cut on invariant mass itself is shown in Fig.~\ref{fig:cdf-trilep} (left). Results were interpreted within mSUGRA.
The exclusion contour in a $(m_0,m_{1/2})$ plane is shown in Fig.~\ref{fig:cdf-trilep} (right).

\begin{figure}[htb]
\centering
\includegraphics[width=0.45\textwidth]{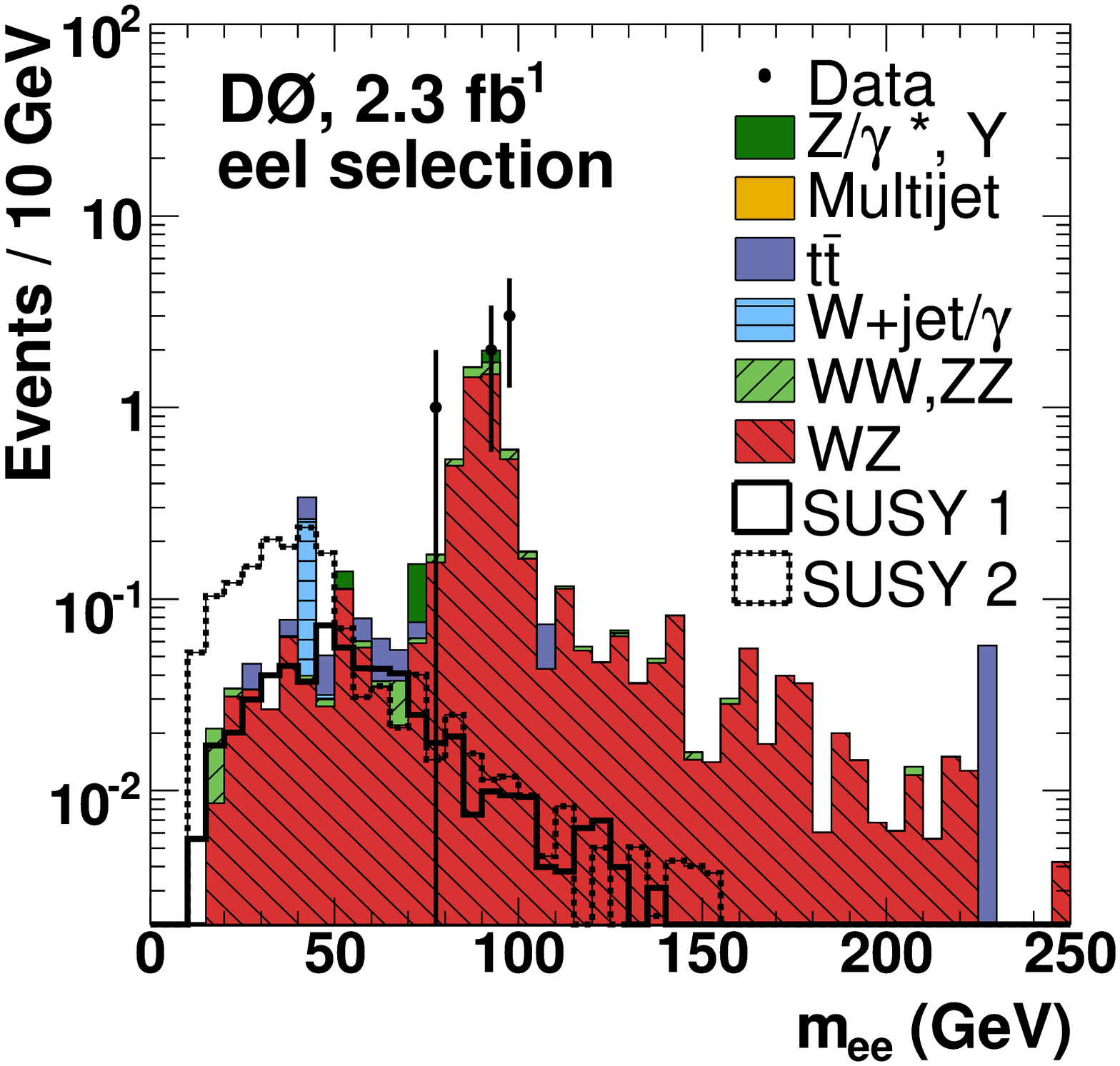}
\includegraphics[width=0.45\textwidth]{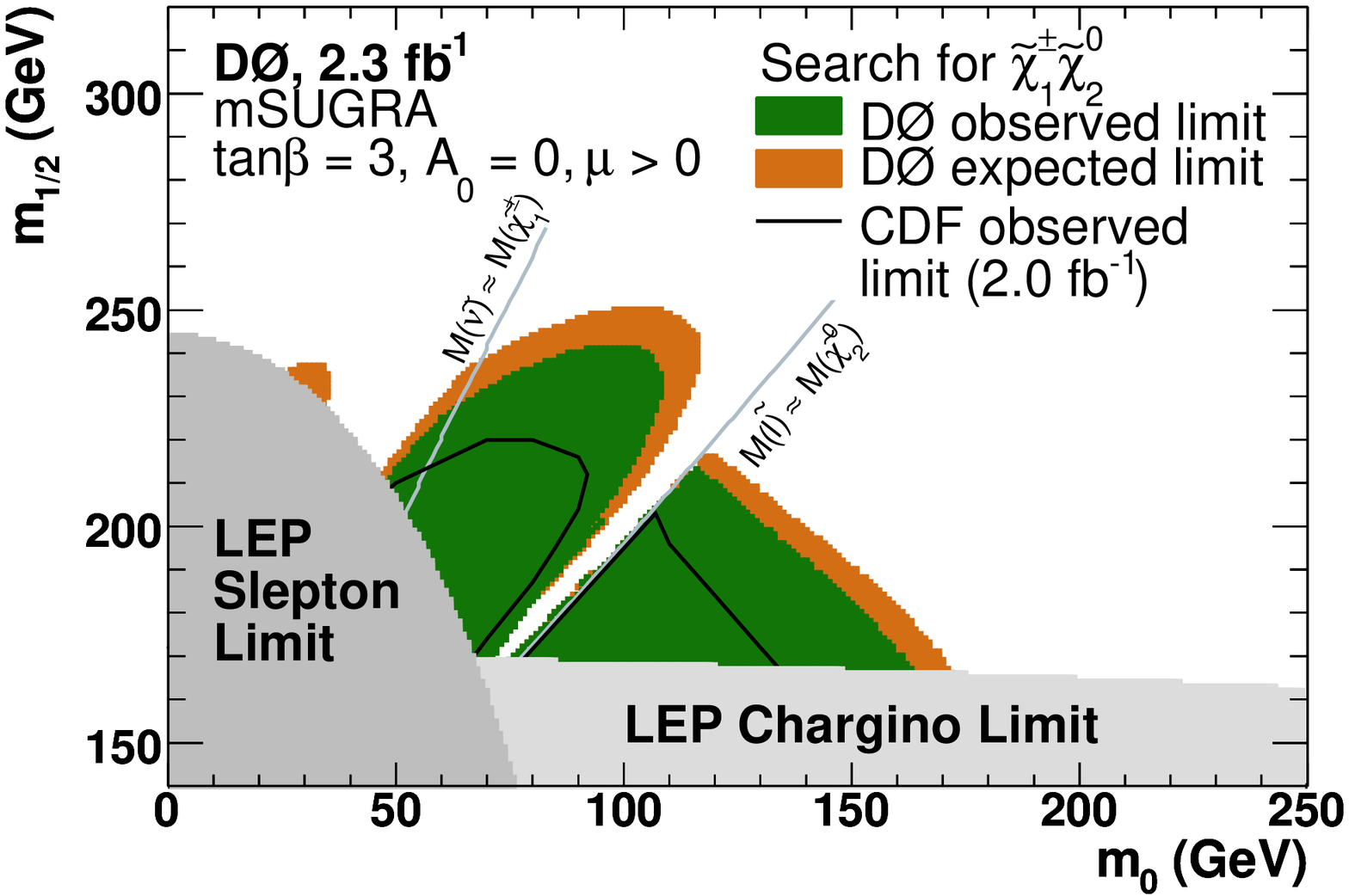}
\caption{Invariant mass of the two electrons after all cuts except the one on $m_{ee}$ for data, expected background and SUSY signal (left). 
The expected and observed 95\% C.L. limit contours in the $(m_0,m_{1/2})$ plane (right).}
\label{fig:d0-trilep}
\end{figure}

D0 defined four different channels distinguished by the lepton content of the final state, dielectron plus lepton ($eel$), dimuon plus lepton ($\mu\mu l$),
electron, muon plus lepton ($e\mu l$), and muon, $\tau$ lepton, where $\tau$ lepton was identified through its hadronic decays, plus lepton. 
For each channel, one "low-$p_T$" and one "high-$p_T$" selection was designed to maximize sensitivity for various parameter points in the $(m_0,m_{1/2})$
plane. Selection requirements included $p_T(l_1)>12$~GeV, $p_T(l_2)>8$~GeV, rejecting events with same flavor dilepton mass in a $Z$ mass window,
 rejecting events with back-to-back leptons, and requiring large $\met$.
 After all selection requirements good agreement between data and SM backgrounds was observed as shown for the invariant mass of two electrons in 
 Fig.~\ref{fig:d0-trilep} (left).  Limits on chargino and neutralino production were set and interpreted in mSUGRA. The excluded region in 
 the $(m_0,m_{1/2})$ plane is shown in Fig.~\ref{fig:d0-trilep} (right).

\begin{figure}[htb]
\centering
\includegraphics[width=0.45\textwidth]{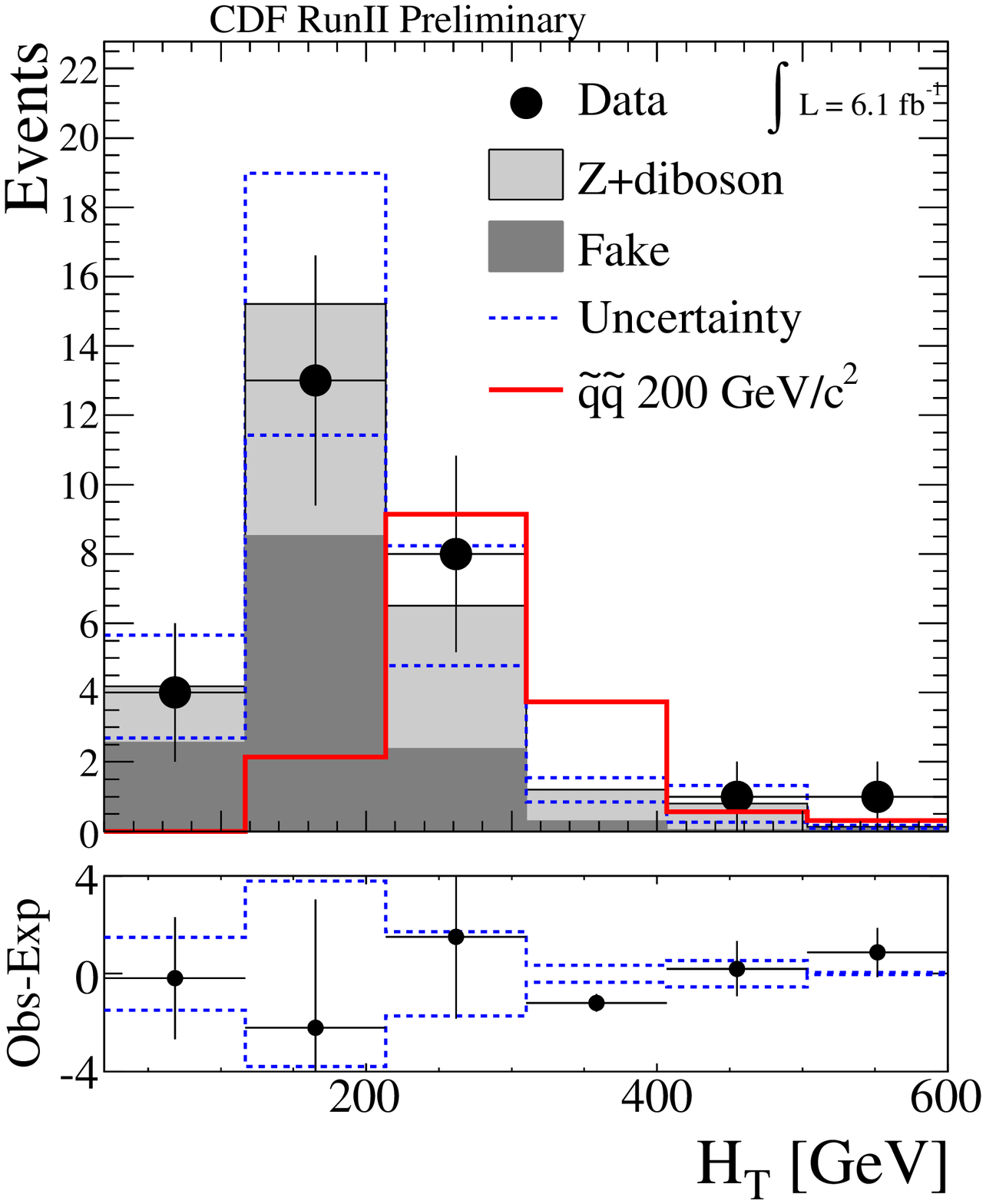}
\includegraphics[width=0.45\textwidth]{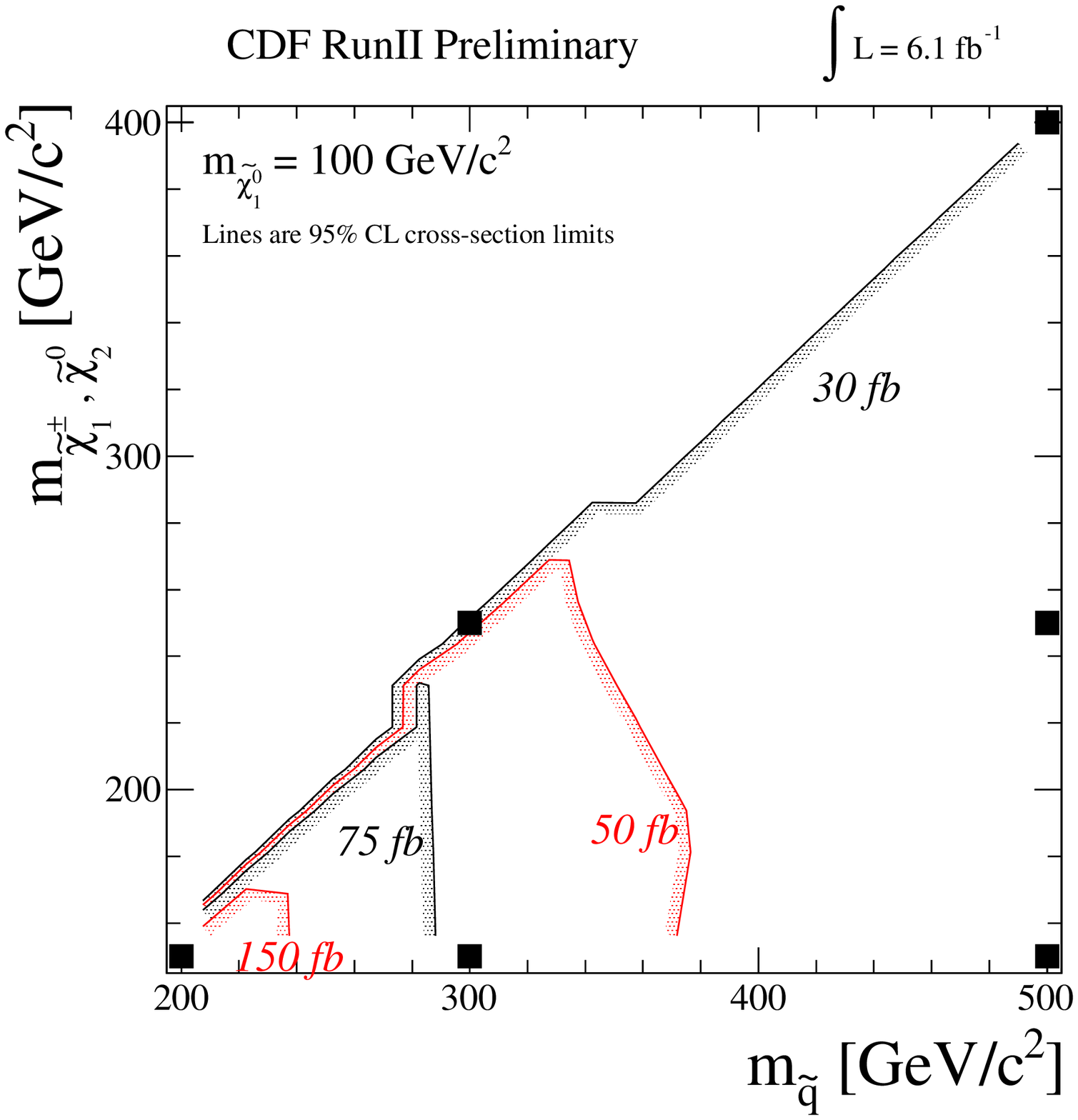}
\caption{The scalar sum of the leptons and jets $p_T$, $H_T$, in the events with the two same sign leptons and at least two jets (left).
95\% C.L. limits on squark pair production for LSP mass of 100 GeV (right). Excluded is the area above the curve.}
\label{fig:cdf-ss}
\end{figure}

CDF searched for SUSY signatures in events with two leptons with the same electric charge~\cite{Aaltonen:samesign} in 6.1~fb$^{-1}$ of data. 
A simplified SUSY model was built, where only particles that 
appear in this model have been tested. This model was based on the following requirements:
\begin{itemize}
\item For final states with leptons, the model contains chargino $\chipm$ and neutralino $\chineu$ which decay to $W$ and $Z$ bosons. 
Slepton modes were not considered.
\item R-parity conservation and the presence of a stable LSP were assumed.
\item The largest cross section will be for pair production of colored states squarks and gluinos.
\end{itemize}
Events were selected with at least two same sign leptons with $p_T>15$~GeV for the 
leading and $p_T>10$~GeV for second leading lepton, and at least two jets with 
$p_T>15$~GeV. The scalar sum of the leptons and jets $p_T$ ($H_T$) is shown in Fig.~\ref{fig:cdf-ss} (left). 
No significant excess in data was observed and 
limits on gluino and squark production were set. 
Figure~\ref{fig:cdf-ss} (right) shows limit on squark pair production for LSP mass of 100 GeV. Excluded area is above the 
curve.

\begin{figure}[htb]
\centering
\includegraphics[width=0.45\textwidth]{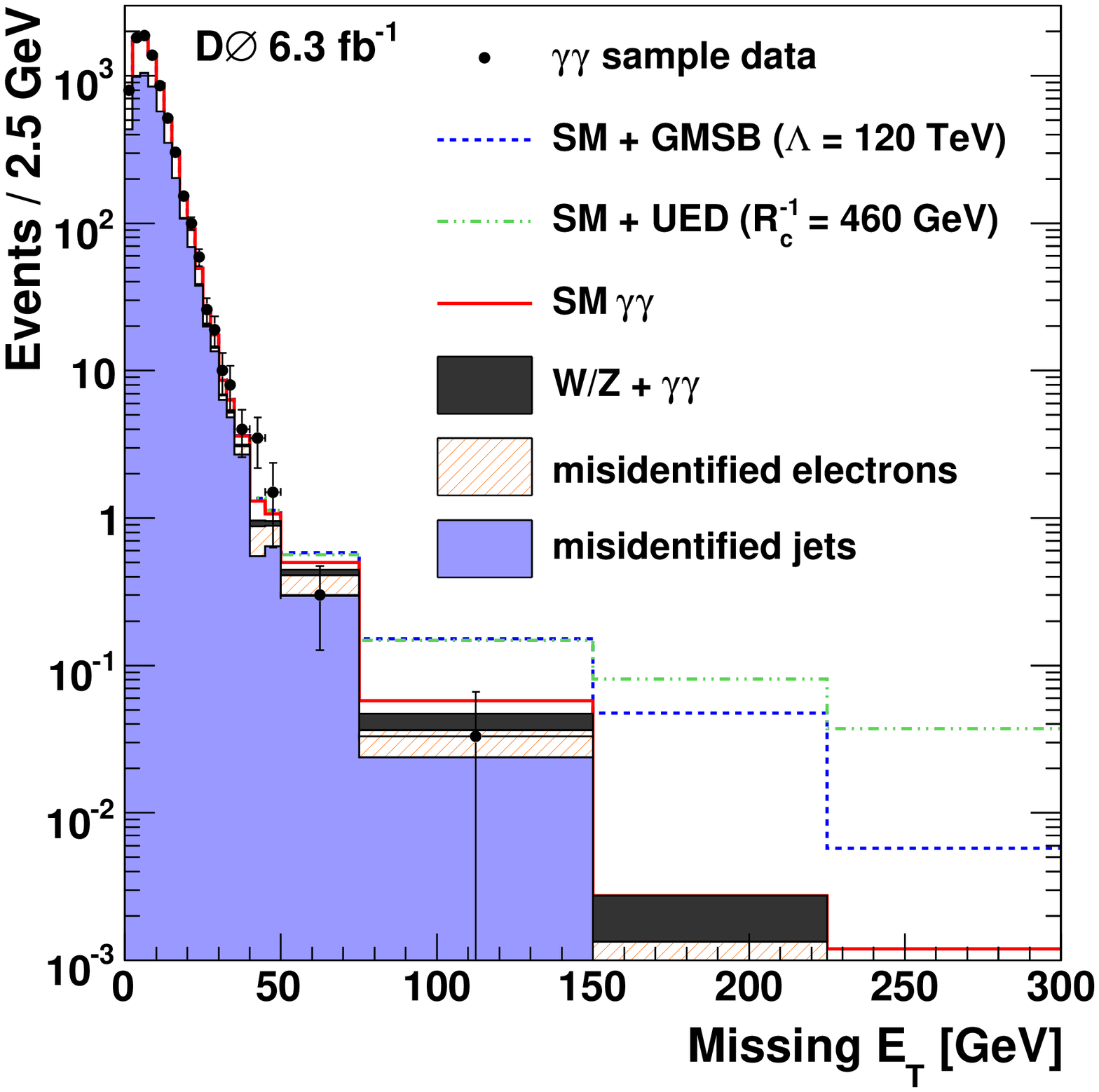}
\includegraphics[width=0.45\textwidth]{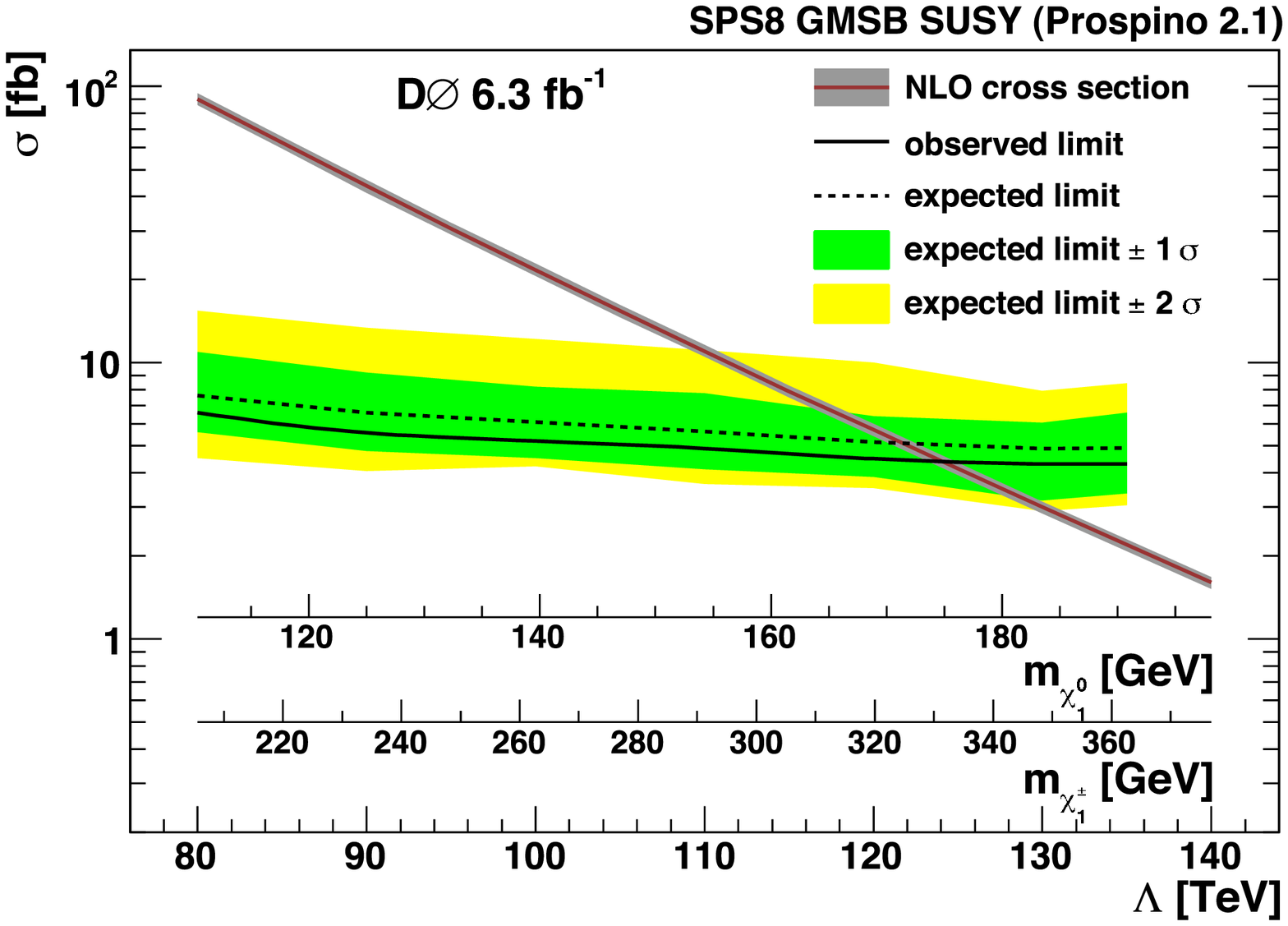}
\caption{$\met$ distribution in $\gamma\gamma$ events for data, predicted backgrounds and expected signal (left). The expected and observed 
95\% C.L. upper cross section limit for GMSB as a function of $\Lambda$, $m_{\chineu}$ and $m_{\chipm}$ (right). }
\label{fig:d0-dipho}
\end{figure}

D0 searched for SUSY in GMSB models
where gravitino $\grav$ is very light and thus is the LSP and the lightest neutralino $\chineu$ is the NLSP~\cite{Abazov:2010us} in 6.3~fb$^{-1}$ of data. 
The largest cross section at Tevatron, assuming R-parity conservation, 
is chargino and neutralino pair production ($\chipm\chineuhi,\chipm\chipm$) with subsequent decay chains to NLSP $\chineu$. A case 
where $\chineu$ decays promptly to photon and an essentially massless gravitino, $\chineu\rightarrow\grav\gamma$, was considered. 
Events were selected with at least two 
isolated photons in central calorimeter region with $p_T>25$~GeV. To remove events with poorly modeled $\met$, events were rejected
 with $\Delta\phi$ between 
$\met$ and leading jet (if present) greater than 2.5, and $\Delta\phi$ between $\met$ and any photon less than 0.2.
SM backgrounds, which were categorized as instrumental $\met$ sources ($\gamma\gamma$, $\gamma$+jet and MJ), and genuine $\met$ sources 
($W\gamma$, $W$+jets, $W/Z+\gamma\gamma$), were estimated from data. Figure~\ref{fig:d0-dipho} (left) shows the $\met$ distribution in 
$\gamma\gamma$ events, where a good agreement between data and SM backgrounds was observed.
A GMSB scenario was probed using the set of parameters from the SPS8 model, where the scale $\Lambda$ was unconstrained, $M_{mes}=2\Lambda$, 
$N_{mes}=1$, $\tan\beta=15$, and $\mu>0$. Since no evidence of the SUSY in the $\met$ distribution was observed, a 95\% C.L. upper limit on the 
cross section production as a function of $\Lambda$, $m_{\chineu}$ and $m_{\chipm}$ was set (Fig.~\ref{fig:d0-dipho} (right)).

\begin{figure}[htb]
\centering
\includegraphics[width=0.45\textwidth]{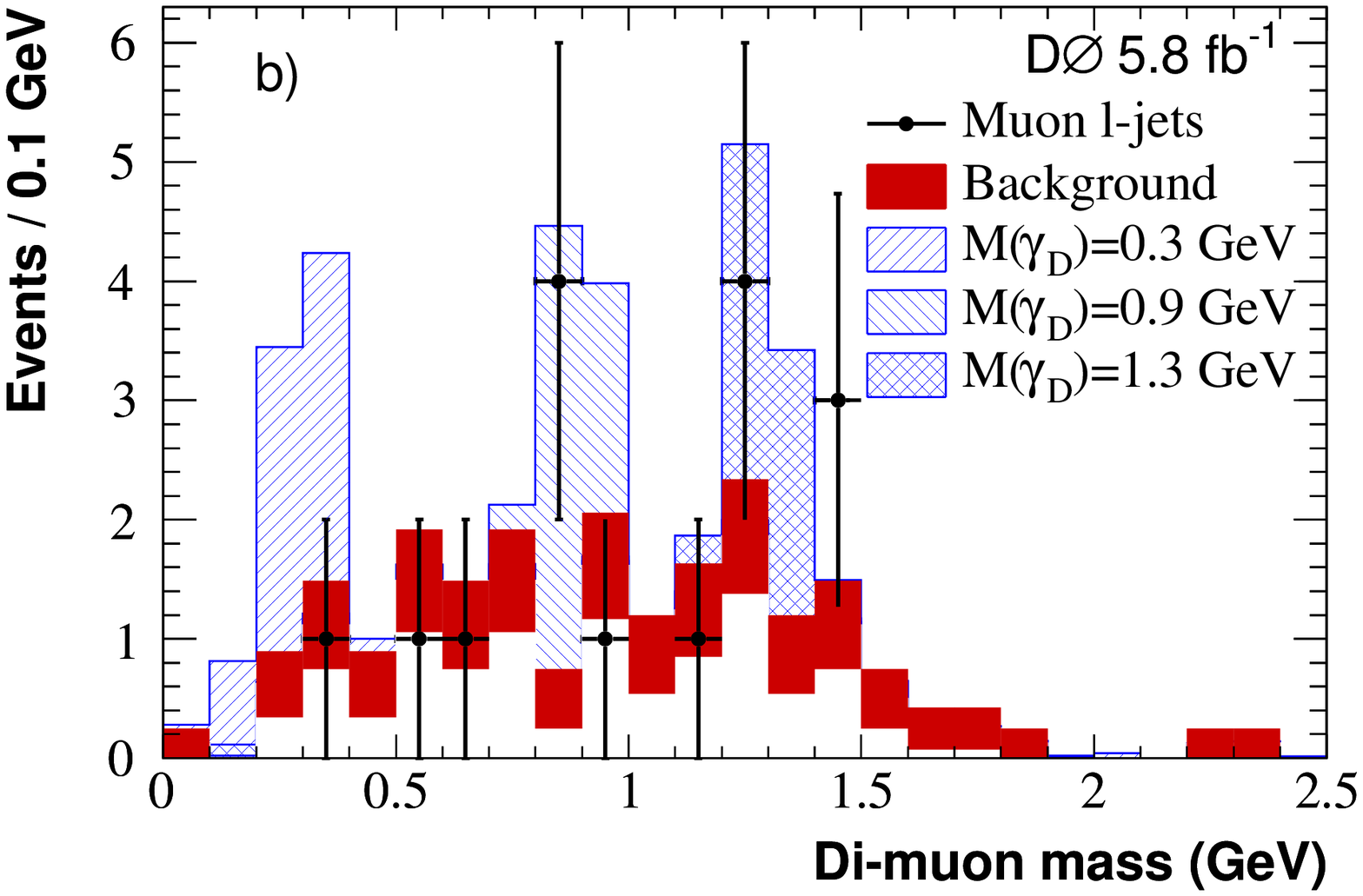}
\includegraphics[width=0.45\textwidth]{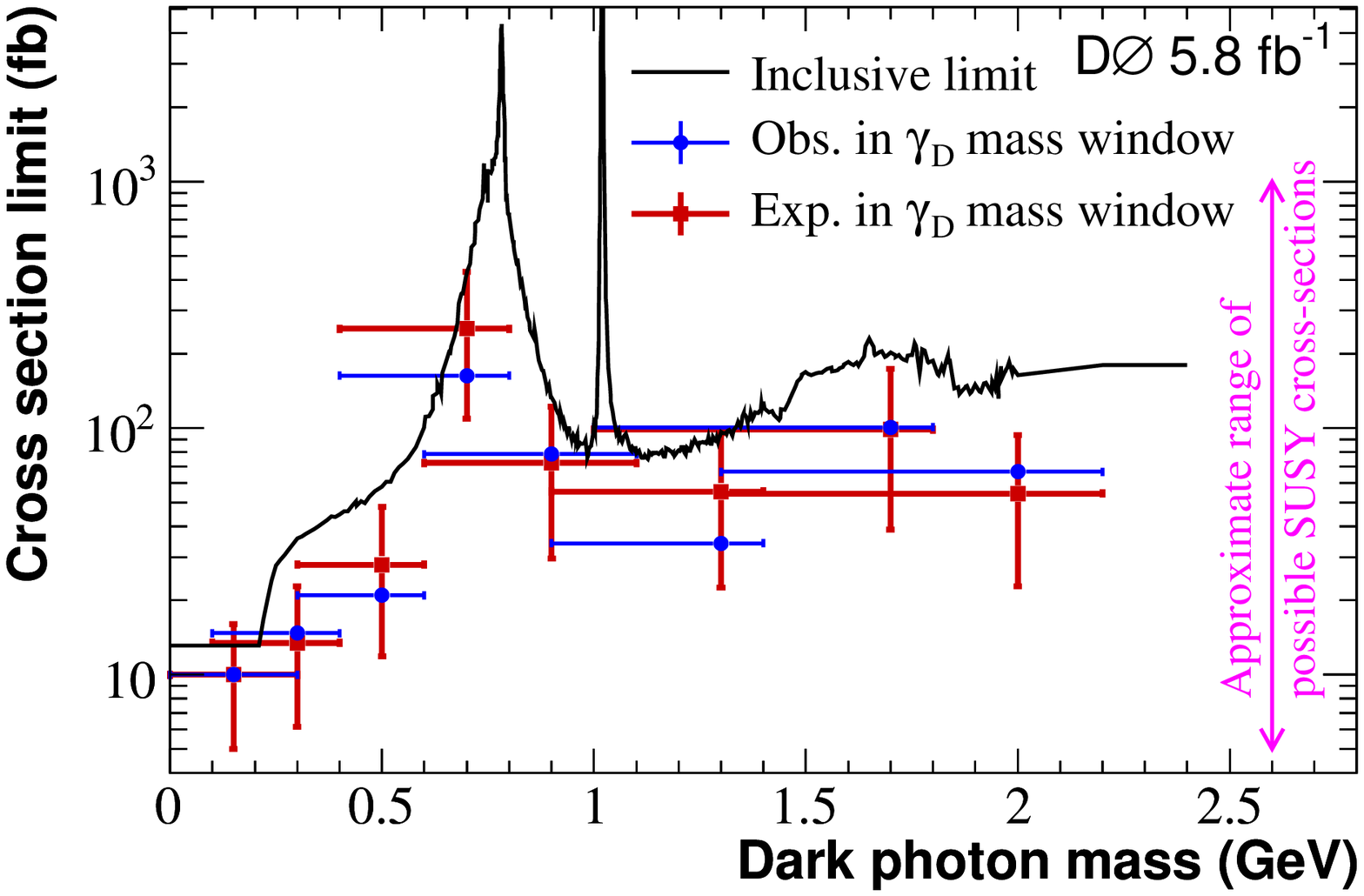}
\caption{Invariant mass of dark photon candidates
with two isolated $l$-jets and $\met>30$~GeV where $\gamma_D$ decayed to two muons (left). 
95\% C.L. limits on the cross section as a function of the $m_{\gamma_D}$ (right). }
\label{fig:d0-lj}
\end{figure}

D0 searched for signatures of hidden-valley models~\cite{Abazov:2010uc} in 5.8~fb$^{-1}$ of data. Hidden valley models contain a hidden sector that
is very weakly coupled to the SM. The force carrier in the hidden-valley model is the dark photon ($\gamma_D$), which is assumed to be
 very light ($m_{\gamma_D}<2$~GeV) and 
decays into a pair of SM charged fermions. In these models, SUSY will have partners for both the SM and the hidden sector. It was 
assumed that the LSP of the hidden 
sector $\tilde{X}$ was lighter than the lightest SM SUSY partner SM-LSP. Then the SM-LSP decayed promptly into particles of the
hidden sector if R-parity was conserved.
Hidden-sector particles are light so their decays produce jets of tightly
collimated particles from decays of $\gamma_D$, i.e. jets of leptons or leptonic jets ($l$-jets).
Since $\gamma_D$ can decay to a pair of electrons or to a pair of muons, $l$-jet can be either "electron $l$-jet" ($e_{lj}$)
or "muon $l$-jet" ($\mu_{lj}$).  Events were selected with two $l$-jets, $e_{lj} e_{lj}$, $e_{lj}\mu_{lj}$ or $\mu_{lj}\mu_{lj}$.
The main backgrounds, which were estimated from a data, consist of MJ production, and, in a case of $e_{lj}$, photon production with a decay to 
$e^+ e^-$ pairs. We searched for presence of leptonic jets in the events with $\met>30$~GeV. Figure~\ref{fig:d0-lj} (left) shows invariant mass of two muon $l$ 
jets for different  assumption of $m_{\gamma_D}$. With no significant excess observed in data, 95\% C.L. upper limits were set as a function of $m_{\gamma_D}$
on the production cross section for SUSY particles decaying
to two $l$-jets and large $\met$ (Fig.~\ref{fig:d0-lj} (right)).


\section{R-parity violation searches}
\begin{figure}[htb]
\centering
\includegraphics[width=0.45\textwidth]{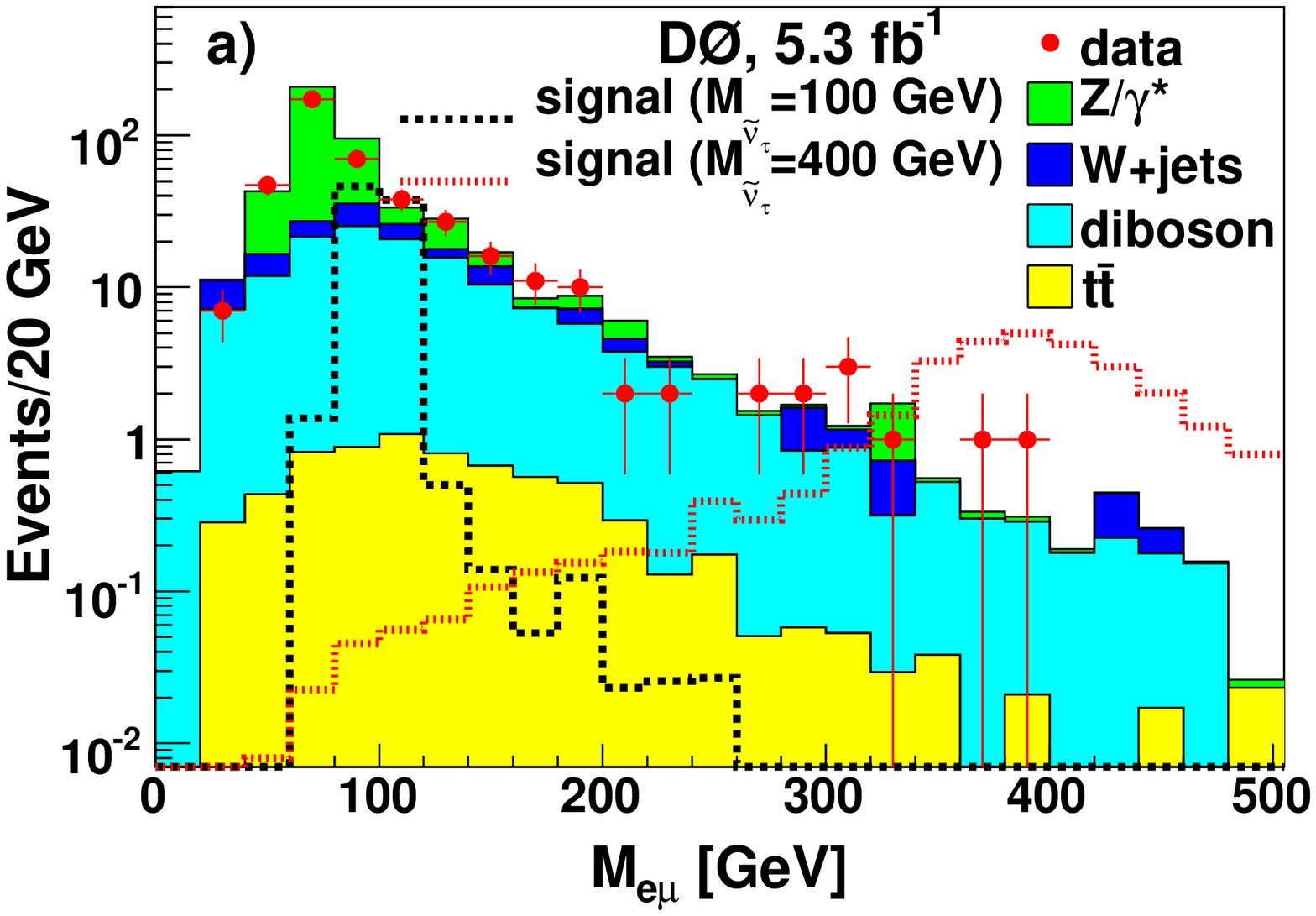}
\includegraphics[width=0.45\textwidth]{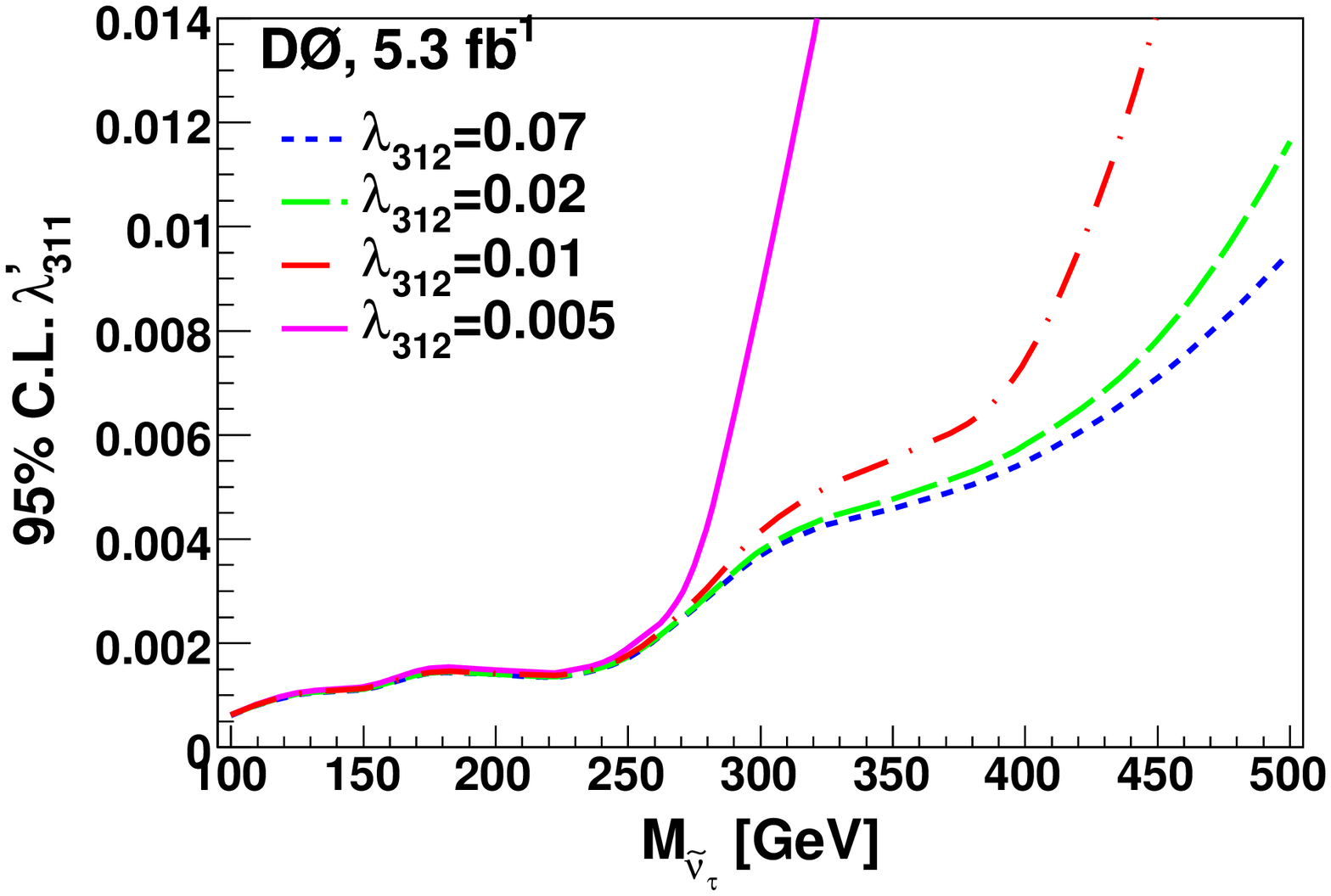}
\caption{Distribution of $M_{e\mu}$ for data, predicted background and expected signal (left). The 95\% C.L. observed upper limits
on $\lambda'_{311}$ for values of $\lambda_{312}$
as a function of $M_{\sneu_{\tau}}$ (right).}
\label{fig:sneu}
\end{figure}

If R-parity is violated single production of SUSY particles is allowed, and they can be observed at Tevatron.
D0 searched for resonant production of a sneutrino which decays into an electron and a muon in 5.4~fb$^{-1}$ of data~\cite{Abazov:2010km}. 
The relevant terms in Lagrangian which describe production and decays of $\sneu$ are:
\begin{equation}
{\cal L}_{RPV}=-\frac{1}{2}\lambda_{ijk}(\sneu_{iL}\bar{l}_{kR} l_{jl}-\sneu_{jL}\bar{l}_{kR} l_{il})
-\lambda'_{ijk}(\sneu_{iL}\bar{d}_{kR} d_{jl})+h.c.
\end{equation}
where $i,j,k=1,2,3$ is fermion generation, $l$ is down type lepton field and $d$ is down type quark field. Thus at Tevatron, single $\sneu$ can be produced
in $d\bar{d}$ scattering. 
It was assumed that only the third 
generation sneutrino ($\sneu_{\tau}$) was produced, and that it was the LSP. 
 Further it is assumed that all the couplings are equal to zero except
$\lambda'_{311}$ and $\lambda_{312}=\lambda_{321}=-\lambda_{231}=-\lambda_{132}$.
The dominant background processes for this search 
are $Z\rightarrow\tau\tau$, $t\bar{t}$, diboson and $W+jets$. Events were selected with one isolated electron with 
$p_T>30$~GeV, one isolated muon with $p_T>25$~GeV and no jets with $p_T>25$~GeV. Signal events have low $\met$, but due to limited 
momentum resolution of the muons, some $\met$ will be present if it is either aligned or anti-aligned with muon. Thus events were rejected 
with $\met>20$~GeV
and $0.7<\Delta\phi(\mu,\met) <2.3$.
Figure~\ref{fig:sneu} (left) shows $M_{e\mu}$ where good agreement between data and SM background was observed. With no significant excess 
present in a data 95\% C.L. upper limits were set on couplings as a function of $M_{\sneu_{\tau}}$, as shown in Fig.~\ref{fig:sneu} (right).

\begin{figure}[htb]
\centering
\includegraphics[width=0.45\textwidth]{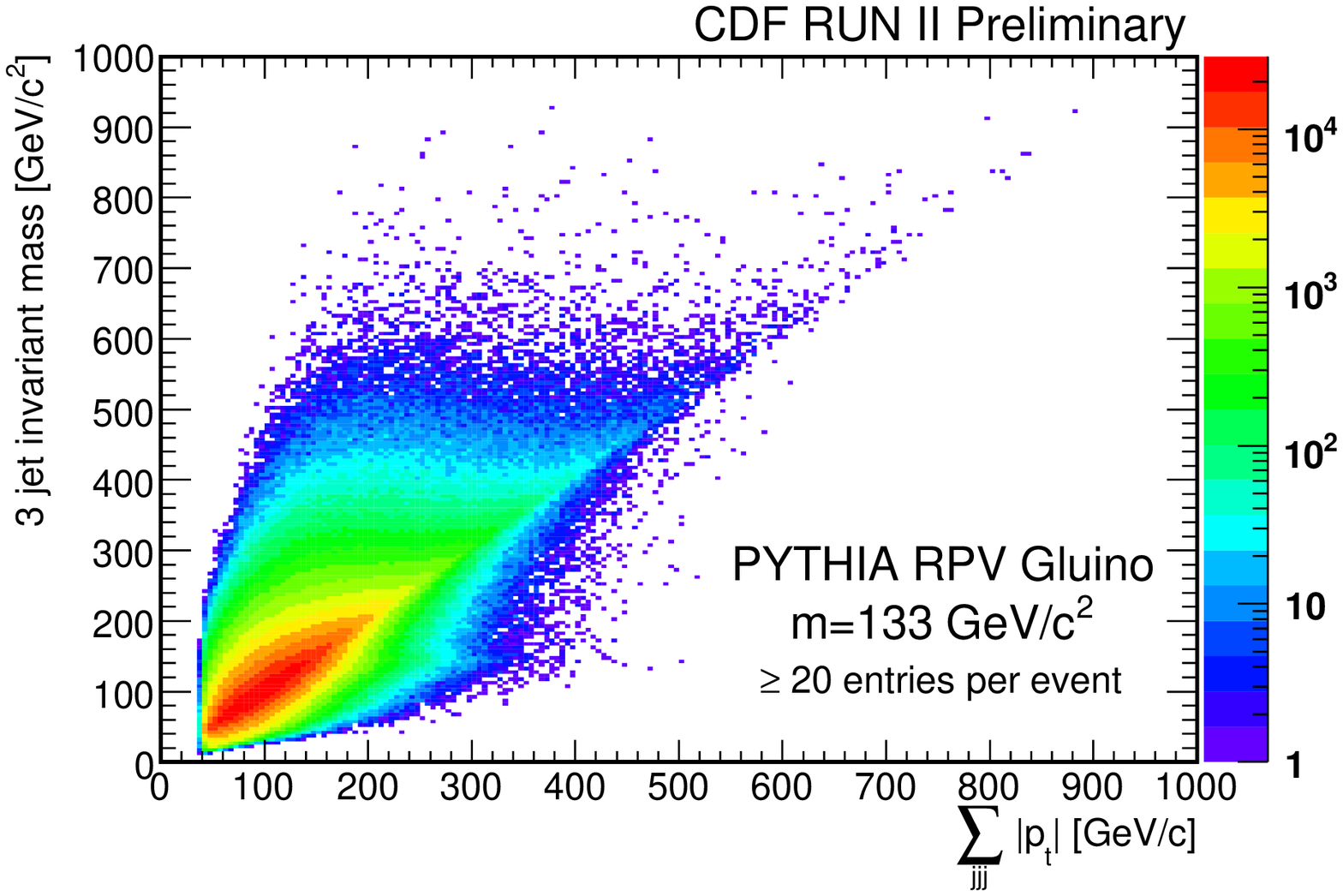}
\includegraphics[width=0.45\textwidth]{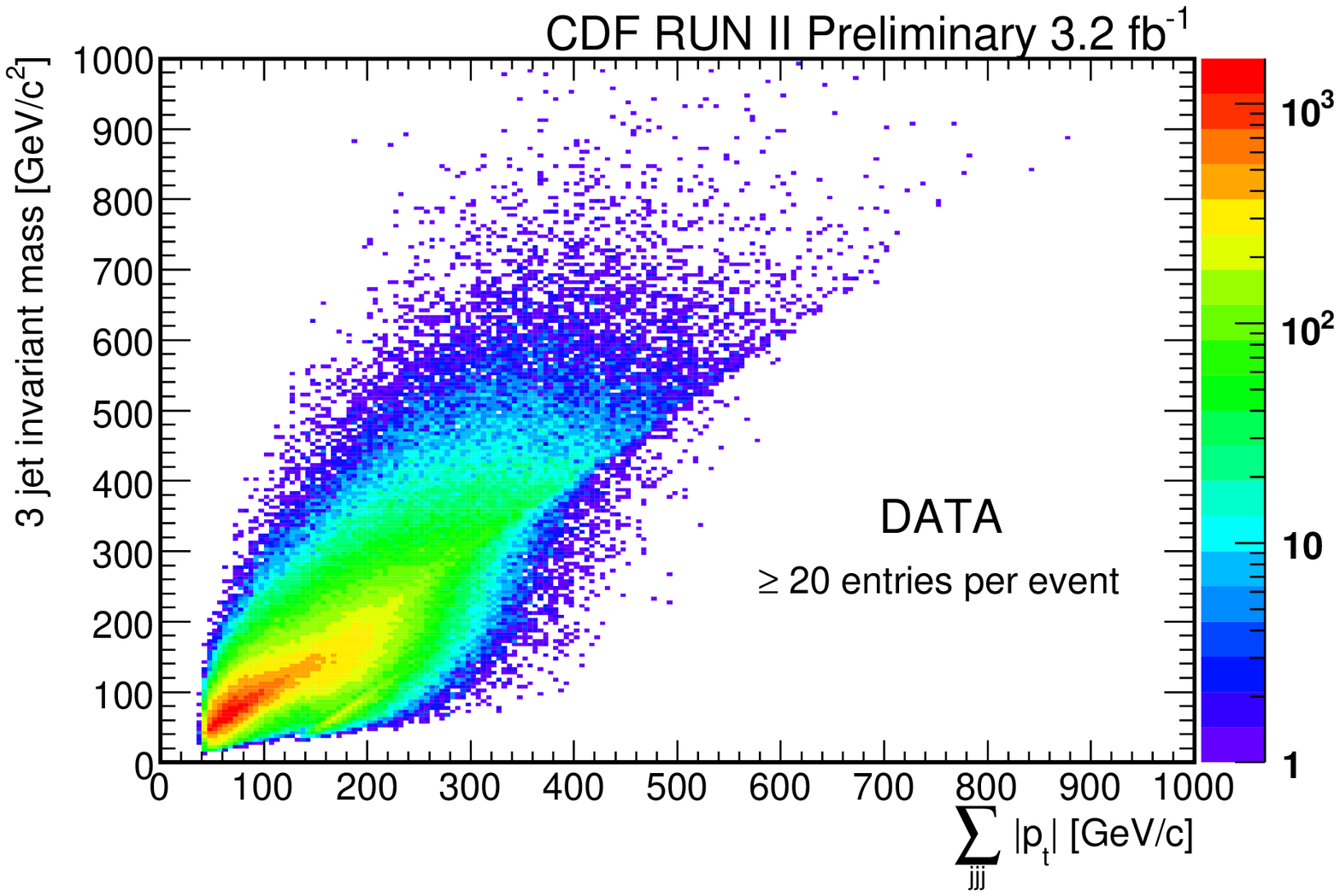}
\caption{Distributions of the $M_{jjj}$ vs $\sum_{jjj}{p_T}$ for the RPV gluino signal (left) and data (right). For each events there are multiple entries ($>20$).}
\label{fig:cdf-3j-2d}
\end{figure}

\begin{figure}[htb]
\centering
\includegraphics[width=0.45\textwidth]{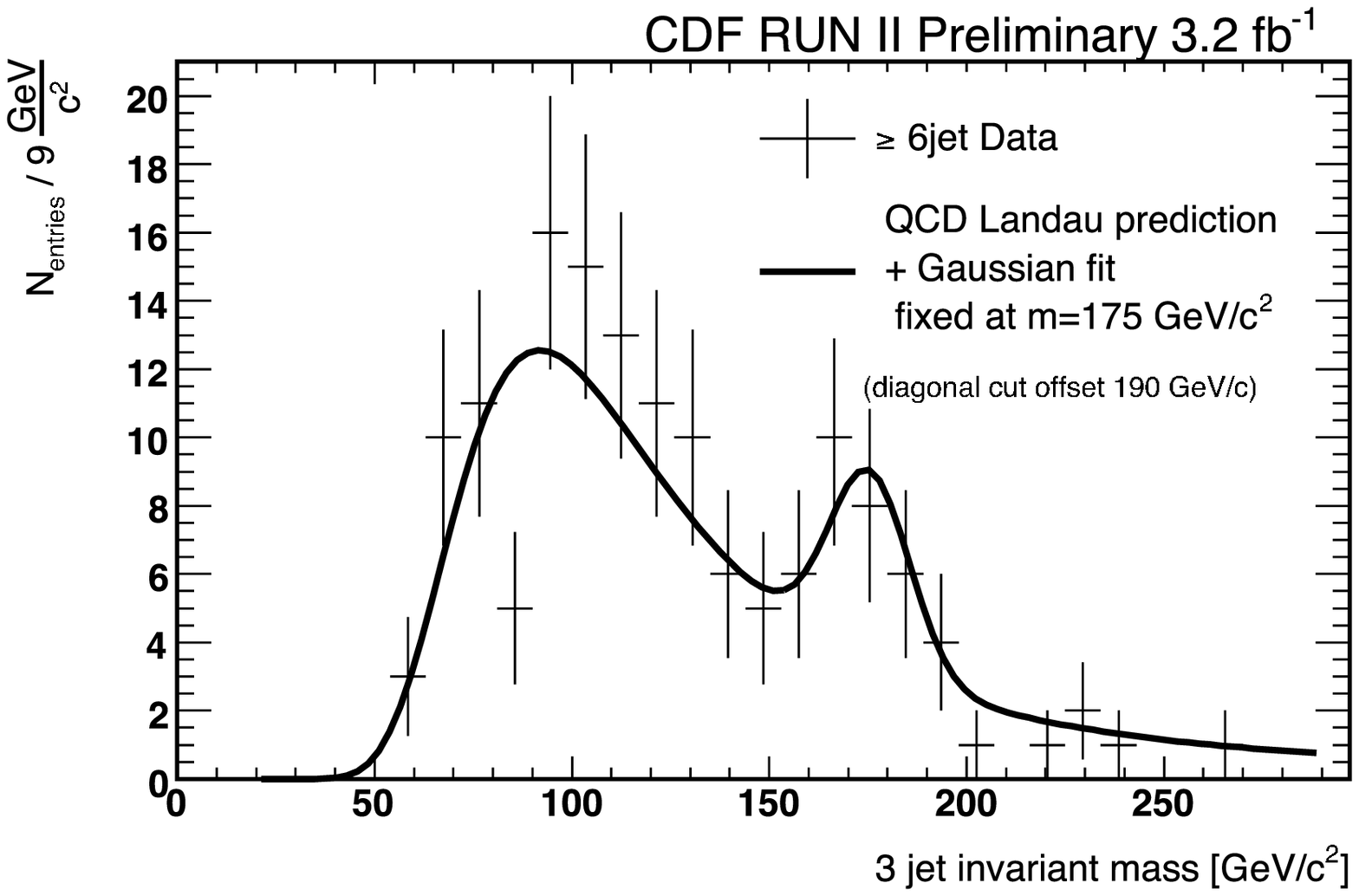}
\includegraphics[width=0.45\textwidth]{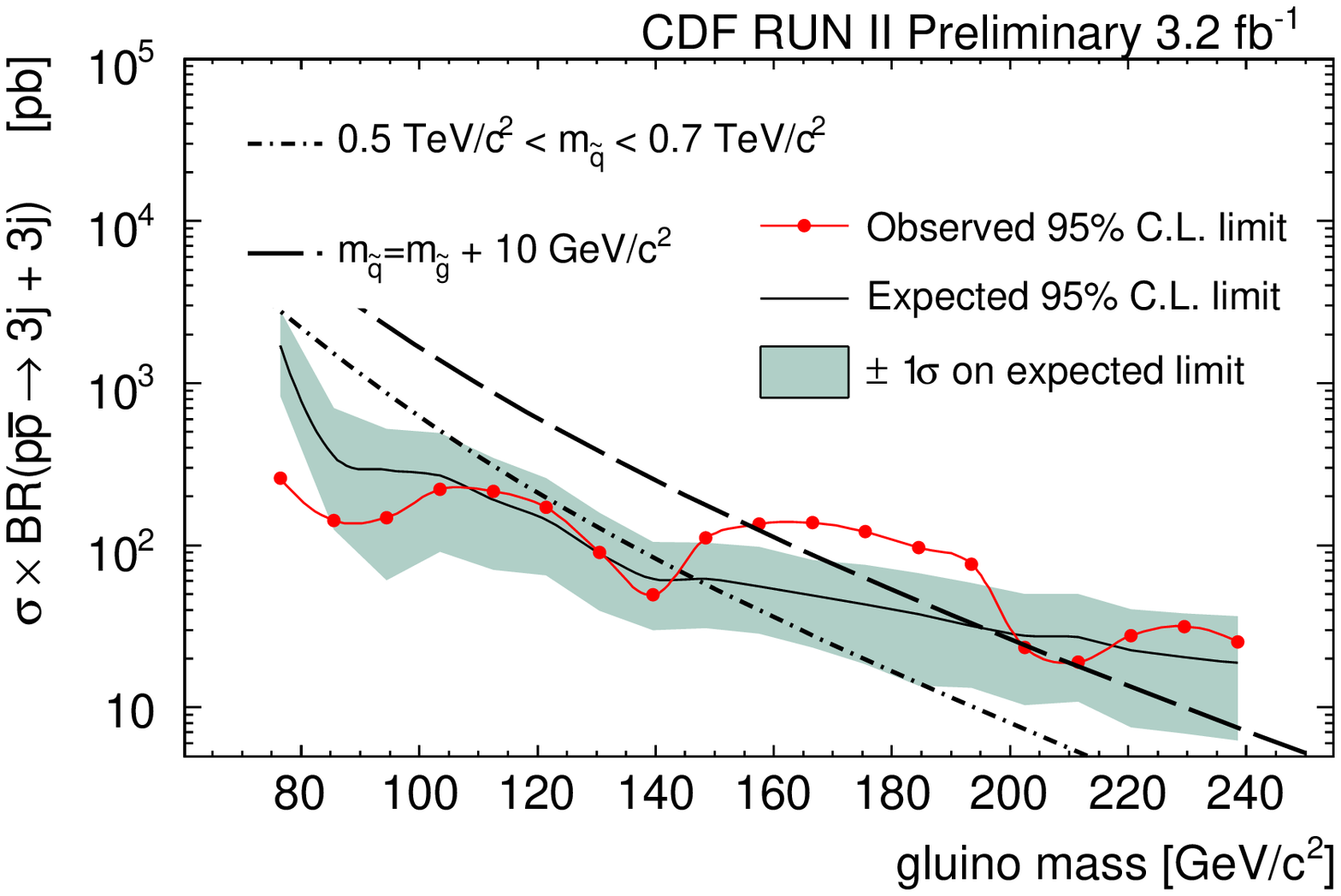}
\caption{Final mass distribution for the data and QCD Landau prediction plus signal Gaussian fit (left). The 95\%
C.L. expected and observed limit (right).}
\label{fig:cdf-3j_res}
\end{figure}

CDF~\cite{Aaltonen:2011sg} searched for 3-jets resonances in 3.2~fb$^{-1}$ of data. 
Although this search was model independent, the possible new physics was modeled with
 RPV SUSY gluino pair production, each gluino decaying into three partons. The dominant background was MJ and it was estimated from data. Events were
 selected with
$\met<50$~GeV, between one and four primary vertices, 
and at least six jets. It was required that the sum of the $p_T$ of the six leading jets was greater than 250 GeV.
The final requirement $\sum_{jjj}{p_T}-M_{jjj}>$~offset, with distributions shown in Fig.~\ref{fig:cdf-3j-2d} (left) for RPV 
gluino with mass of 190 GeV and Fig.~\ref{fig:cdf-3j-2d} (right) for data,
was optimized for different mass points. MJ background was estimated from five jets data and fitted with Landau function.
To extract signal from combinatorial background Landau+Gaussian fit was used (see Fig.~\ref{fig:cdf-3j_res} (left)), and number of signal events 
was obtained by integrating the Gaussian in $\pm 1\sigma$ range.
No significant excess 
was observed in data, and a 95\% C.L. limit on the cross section times branching ratio as a function of gluino mass was set, and it is shown in 
Fig.~\ref{fig:cdf-3j_res} (right).

\section{Conclusion}
Results from SUSY searches at the Tevatron have been presented. It is shown that data agrees with SM expectations, and thus SUSY parameter 
space has been constrained. Limits set by various CDF and D0 searches are the most stringent to date.

\Acknowledgements
D0 and CDF thank the staffs at Fermilab and collaborating institutions.
D0 acknowledge support from the
DOE and NSF (USA);
CEA and CNRS/IN2P3 (France);
FASI, Rosatom and RFBR (Russia);
CNPq, FAPERJ, FAPESP and FUNDUNESP (Brazil);
DAE and DST (India);
Colciencias (Colombia);
CONACyT (Mexico);
KRF and KOSEF (Korea);
CONICET and UBACyT (Argentina);
FOM (The Netherlands);
STFC and the Royal Society (United Kingdom);
MSMT and GACR (Czech Republic);
CRC Program and NSERC (Canada);
BMBF and DFG (Germany);
SFI (Ireland);
The Swedish Research Council (Sweden);
and
CAS and CNSF (China).
CDF acknowledge support from the
DOE and NSF (USA);
INFN (Italy);
Ministry of Education, Culture, Sports, Science and Technology (Japan);
NSERC (Canada); 
National Science Council (China);
National Science Foundation (Switzerland);
A.P. Sloan Foundation (USA); 
BMBF (Germany); 
World Class University Program and National Research Foundation (Korea);
STFC and the Royal Society (United Kingdom);
CNRS/IN2P3 (France);
RFBR (Russia);
Ministerio de Ciencia e Innovacion, and Programa Consolider-Ingenio 2010 (Spain);
R\&D Agency (Slovakia);
Academy of Finland; 
and the Australian Research
Council (ARC).


\begin{thebibliography}{99}
\bibitem{eta}
The pseudorapidity is defined as $\eta=-\ln[\tan(\theta/2)]$, where $\theta$ is
  the polar angle with respect to the proton beam direction.
\end{thebibliography}
\end{document}